\documentclass[twocolumn, trackchanges, twocolappendix]{aastex701}

\usepackage{physics}
\usepackage[dvipsnames]{xcolor}

\begin{document}

\shorttitle{Evolution of Low-Mass Pop~III Stars} \shortauthors{T. Ferreira et al.}

\title{Evolution of Low-Mass Population III Stars: \\ Convection, Mass Loss, Nucleosynthesis, and Neutrinos}

\correspondingauthor{Thiago Ferreira dos Santos}
\author[orcid=0000-0003-2059-470X,gname='Thiago',sname='Ferreira dos Santos']{Thiago Ferreira}
\affiliation{Department of Astronomy, Yale University, 219 Prospect St., New Haven, CT 06511, USA}
\email[show]{thiago.dossantos@yale.edu}  

\author[orcid=0000-0003-4456-4863,sname='Bellinger']{Earl P. Bellinger}
\affiliation{Department of Astronomy, Yale University, 219 Prospect St., New Haven, CT 06511, USA}
\email[]{earl.bellinger@yale.edu}

\author[orcid=0000-0002-5794-4286,sname='Farag']{Ebraheem Farag}
\affiliation{Department of Astronomy, Yale University, 219 Prospect St., New Haven, CT 06511, USA}
\email[]{ebraheem.farag@yale.edu}

\author[orcid=0000-0001-8722-1436,sname='Lindsay']{Christopher J. Lindsay}
\affiliation{Department of Astronomy, Yale University, 219 Prospect St., New Haven, CT 06511, USA}
\email[]{christopher.lindsay@yale.edu}

\begin{abstract}

    The first stars likely formed from pristine clouds, marking a transformative epoch after the dark ages by initiating reionisation and synthesising the first heavy elements. Among these, low-mass Population III stars are of particular interest, as their long lifespans raise the possibility that some may survive to the present day in the Milky Way's stellar halo or satellite dwarfs. As the first paper in a series, we present hydrodynamic evolutionary models for $0.7-1~M_\odot$ stars evolved up to the white dwarf phase, utilising the MESA software instrument. We systematically vary mass-loss efficiencies, convective transport, and overshooting prescriptions, thereby mapping how uncertain physics influences nucleosynthetic yields; surface enrichment, including nitrogen-rich post-main sequence stars arising from convective shell mergers; remnant properties, such as low-mass helium or carbon-oxygen white dwarfs ($M_{\rm WD} \sim 0.45-0.55~M_\odot$) and transient UV-bright phases; and potential observational signatures, including neutrino emission during shell mergers and helium flashes. These models establish a predictive framework for identifying surviving Pop~III stars and their descendants, providing both evolutionary and observational constraints that were previously unexplored.

\end{abstract}


\section{Introduction}\label{sec:introduction}

Within the standard Lambda Cold Dark Matter ($\Lambda{\rm CDM}$) model, primordial gas cools via collisional excitation of atomic hydrogen, helium, and singly ionised helium, allowing it to reach temperatures of $\sim$200 K \citep{1967Natur.216..976S, 1968ApJ...154..891P, 1969PThPh..42..219M, 1981MNRAS.197.1021C, 1983MmSAI..54..235P, 1986PASP...98.1081S, 2000ApJ...540...39A, 2012MNRAS.426..377G}, which is significantly higher than the $\sim$10 K characteristic of present-day star-forming gas \citep{2001ApJ...546..635O, 2018ApJ...862..102S}. Since accretion rates in dark matter mini-haloes scale with temperature as $\propto T^{3/2}$, the first stars formed in the Universe (henceforth Population III; Pop~III stars or primordial stars) were initially predicted to be exceptionally massive ($\sim$100–300 $M_\odot$; \citealt{1976MNRAS.176..483R, 1999ApJ...527L...5B, 2004ARA&A..42...79B}). However, in regions with significant levels of ultraviolet (UV) radiation capable of dissociating molecular hydrogen (often called Lyman-Werner (LW) photons with energies $11.2~{\rm eV} \lesssim h\nu < 13.6~{\rm eV}$; \citealt{2024ApJ...965..141N}), the cooling and collapse of molecular clouds are delayed, leading to higher accretion rates and efficient disc fragmentation. This favours not only the formation of multiple star systems but also a bimodal mass distribution for Pop~III stars, where one mode favours high-mass stars, whilst another allows for the formation of low-mass stars (down to $\sim$0.8 $M_\odot$) if their formation is accompanied by ejection via three-body interactions (e.g., scattering or von Zeipel–Lidov–Kozai oscillations; \citealt{1996ApJ...473L..95U, 1999ApJ...515..239N, 2014ApJ...785...73S, 2014ApJ...792...32S, 2015MNRAS.448..568H, 2016ARA&A..54..441N, 2020ApJ...901...16D}; see also \citealt{2019MEEP....7....1I} for in-depth discussions about the mechanism in particular). Such low-mass survivors could still be present in the Milky Way and/or in its dwarf satellites \citep{2016ApJ...820...59K, 2017MNRAS.470..898H, 2018ApJ...867...98S, 2018MNRAS.473.5308M, 2020ApJ...901...16D, 2021AJ....161..197C, 2021MNRAS.503.6026R}, possibly identified as very/extremely metal-poor red giant stars or remnants such as white dwarfs (WDs) for the highest masses \citep{2002RvMP...74.1015W, 2004A&A...416.1117C, 2008ApJ...681.1524L, 2010ApJ...724..341H}.

Given their pristine composition---hydrogen, helium, and trace lithium---Pop~III stars are expected to exhibit distinct internal structures compared to their metal-rich counterparts. For instance, lacking heavy elements and possessing higher surface and central temperatures, they are expected to have significantly lower opacities, allowing radiation to escape more efficiently. This results in more compact structures to maintain hydrostatic equilibrium, increased energy generation rates, and higher luminosities. Consequently, lower metallicity stars appear hotter (bluer) and more compact on the Hertzsprung–Russell (H–R) diagram for a certain mass. Their main sequence lifetimes are also likely shorter due to reduced opacity-driven radiation pressure \citep{2004sipp.book.....H, 2013sse..book.....K}. In addition, the absence of metals severely curtails radiation-driven winds, enabling even the most massive Pop III stars to retain a significant portion of their mass until advanced helium-burning phases, profoundly influencing their evolutionary endpoints and nucleosynthetic yields \citep{1990ApJ...349..580F}. 

Mass remains the dominant factor in determining the fate of Pop~III stars, spanning a broad range from low-mass stars that undergo helium flash at the tip of the red giant branch (TRGB) to ultra-massive stars prone to relativistic collapse (see e.g., \citealt{2001A&A...371..152M, 2003A&A...399..617M, 2025arXiv251011772S, 2025arXiv251017952S}). Pop~III stars with $M < 0.8~M_\odot$ could still be on the main sequence today, given that their lifetimes exceed the Hubble time; however, unlike evolved giants that may be detected in ultra-faint dwarf galaxies (e.g., \citealt{2018MNRAS.473.5308M, 2021MNRAS.503.6026R}), these low-mass stars remain intrinsically faint, posing a major challenge for detection with current telescopes. Low-mass Pop~III stars ($0.8 < M \lesssim 1.1~M_\odot$) ignite helium in degenerate cores and are expected to undergo helium flash episodes. Due to their lack of metals, these stars exhibit truncated RGB evolution, and some may instead follow extended horizontal branch (EHB) tracks, particularly for masses around $0.85$–$0.9~M_\odot$ \citep{1996ApJ...466..359D, 2023MNRAS.525.4700L}, owing primarily to enhanced mass loss along the RGB ascent; the consequences of which we examine in detail in the next sections. The lowest-mass ($\sim 0.8$–$1.2~M_\odot$) Pop~III stars are predicted to evolve into WDs \citep{2021AJ....161..197C, 2023MNRAS.525.4700L, 2024MNRAS.534.1561D}, whilst slightly more massive counterparts ($\approx1.5-3~M_\odot$) may end their lives as weak or fallback supernovae \citep{2003Natur.422..871U, 2013ARA&A..51..457N, 2014ApJ...785...98T, 2016ApJ...833...20Y, 2017MNRAS.467.4731C}.

Although surface pollution and observational uncertainties can obscure their origins, the internal structure of the first stars is a valuable window into early stellar physics. In this work, as the first of a series, we explore the structure and evolution of metal-free stars with initial masses between $M_{\rm ZAMS} = 0.7$ and $1.0~M_\odot$. 
{Whilst the evolution of such stars has been studied extensively, much of the earlier literature predates modern microphysical treatments, lacks hydrodynamical context, or does not include the final known evolutionary stages. We highlight the persistence of trace hydrogen in the cores of low-mass Pop III stars close to the main-sequence turn-off/terminal-age main sequence (MSTO/TAMS), which allows the production of $^{12}$C by the $3-\alpha$ process to trigger a CNO-cycle-driven hydrogen flash near the end of hydrogen burning, as first found by \cite{1983A&A...118..262G}. Subsequent work examined helium flashes and the dredge-up of CN-cycle-processed material along the red giant branch in detail \citep{1990ApJ...349..580F, 1993ApJS...88..509C, 2004ApJ...609.1035P}. The evolution of low-mass primordial stars up to the tip of the red giant branch (T-RGB) was further explored by \cite{2007ApJ...667.1206S}, who found a critical mass of ($<1.2~M_\odot$) for the onset of off-centre helium ignition. \cite{2008A&A...490..769C} explored nucleosynthetic yields of low- and intermediate-mass Pop III stars for comparison with observed abundances of extremely metal-poor halo stars, introducing the dual/double core flash (DCF) to describe proton-mixing events occurring during the helium core flash following the RGB phase. And more recently, the first full evolutionary calculations following low-mass, zero-metallicity stars through the white dwarf stage were presented by \cite{2023MNRAS.525.4700L}, establishing the feasibility of primordial low-mass white dwarfs and providing baseline predictions for their remnant masses, internal structure, and cooling behaviour.} {Here, we extend this body of literature by investigating the sensitivity of primordial low-mass stellar evolution to uncertainties in stellar physics, in particular the treatment of convective efficiency, mass-loss prescriptions, and the inclusion of hydrodynamic evolution.}

Non-radial oscillation diagnostics are deferred to Paper II (Ferreira et al., {\it submitted}), which builds on the present models to study oscillation spectra and seismic properties that may distinguish Pop~III stars from metal-rich populations beyond spectroscopy. 

In \S\ref{sec:inputphysics}, we outline the physical assumptions adopted in this work. The evolutionary tracks analysed focus primarily on the $0.85~M_\odot$ Pop~III model, which serves as the central case study throughout the paper, as well as in Paper II. Full details for other stellar masses in the $0.7-1~M_\odot$ range are provided in Appendix \ref{app:evolution}. We explore the impact of varying metallicities in \S\ref{sec:comparison_highZ}, and investigate the sensitivity of the $0.85~M_\odot$ model to variations in mass loss efficiency during the red giant branch (\S\ref{sec:mass-loss}), mixing length (\S\ref{sec:MLT}), semi-convection and overshooting (\S\ref{sec:semiconv}). We briefly examine the time-dependent neutrino emission from low-mass Pop~III stars in \S\ref{sec:neutrinos}, highlighting the distinctive bursts and spectral signatures predicted for the $0.85~M_\odot$ model during shell mergers and helium flashes that may offer alternative detection strategies for such faint stars beyond electromagnetic observations. Our discussion and conclusions are summarised in \S\ref{sec:conclusions}.

\section{Input Physics}\label{sec:inputphysics}

Stellar models were computed using the Modules for Experiments in Stellar Astrophysics software instrument ({\sc MESA} {\tt r24.08.1}; \citealt{2011ApJS..192....3P, 2013ApJS..208....4P, 2015ApJS..220...15P, 2018ApJS..234...34P, 2023ApJS..265...15J}) assuming primordial composition with $Z = 0$, and initial mass fractions $X = 0.7551$ and $Y = 0.2449$ \citep{2020A&A...641A...6P} for Pop~III cases\footnote{{MESA {\tt inlists} can be found at \\ \url{https://github.com/thiagofst/PopIII}}}. Masses ranged from $0.7$ to $1$ $M_\odot$ in $0.05~M_\odot$ increments. We also evolved ultra-low and solar-metallicity models ($Z = 10^{-10}, ~10^{-7}, ~10^{-5}, ~10^{-3}$ and $10^{-2}$) for comparisons in \S\ref{sec:comparison_highZ} and Paper II. We evolved these stellar models from the pre-main sequence to the white dwarf (WD) phase, neglecting effects of rotation and magnetic fields. 

Big Bang nucleosynthesis (BBN) also produces small quantities of deuterium, $^3$He, and $^7$Li in addition to hydrogen and helium. Among these, lithium is particularly fragile, being efficiently destroyed in stellar interiors through the reactions $^6{\rm Li}(p,\alpha)^3{\rm He}$ and $^7{\rm Li}(p,\alpha)^4{\rm He}$ once temperatures exceed approximately $2.5\times10^6~{\rm K}$ \citep{1948Natur.162..680G, 1948PhRv...73..803A, 1957RvMP...29..547B, 1966PhRvL..16..410P, 1967ApJ...148....3W, 1982A&A...115..357S}. During the pre-main-sequence contraction and subsequent main-sequence evolution of Population~III stars, central temperatures rapidly surpass this threshold, leading to the almost complete depletion of primordial lithium. {Owing to its vanishingly small residual abundance and rapid destruction, lithium exerts essentially no measurable influence on stellar structure, energy generation, or opacity, and is therefore explicitly set to zero in the initial abundances of our models. The production of lithium is, however, possible through our adopted nuclear network at earlier stages of evolution.} 

\noindent {\it Opacity}---Radiative opacities were computed using OPAL Type-1 and Type-2 tables \citep{1993ApJ...412..752I, 1996ApJ...464..943I} for high temperatures ($T \gtrsim 10^4~K$). For low temperatures, opacity data from \cite{2005ApJ...623..585F} were employed. Under degenerate regimes, we used conducive opacities from \cite{2007ApJ...661.1094C}. 

\noindent {\it Equation of State (EoS)}---We used SCvH/OPAL tables for H/He mixtures and hot, dense plasma \citep{1995ApJS...99..713S, 2002ApJ...576.1064R}, FreeEOS for non-solar compositions \citep{2012ascl.soft11002I}, HELM tables for fully ionized plasma \citep{2000ApJS..126..501T}, and Skye for electron-degenerate matter in extreme conditions \citep{2021ApJ...913...72J}. 

\noindent {\it Mass Loss and Atmosphere}---Mass loss on the RGB follows the Reimers' prescription \citep{1977A&A....61..217R} with $\eta_{\rm R} = 0.1,~0.5$ (dubbed {\it canonical} model and applied for all masses considered in this work)$,~0.7$ and $0.9$ (see \S\ref{sec:mass-loss}), whilst the post-RGB phase includes enhanced dust-driven winds using Bl{\"o}cker's prescription \citep{1995A&A...299..755B} with efficiencies of $\eta_{\rm Bl.} = 0.1$. Outer boundary conditions adopt the Eddington ${\rm T}-\tau$ relation \citep{1926ics..book.....E} with an external pressure factor of $P_{\rm ext.} = 1.5$ to better capture radiative effects in optically thin layers (e.g., \citealt{1952PASJ....4...91M, 2023ApJS..265...15J}). {Radial pulsations were suppressed during late post-RGB phases via hydrodynamic drag in the envelope, as in \citet{2024ApJS..270....5F}, to maintain numerical stability during highly luminous, low-envelope-mass stages, where the implicit hydrodynamical schemes would otherwise attempt to resolve short-timescale dynamical responses beyond the scope of this work.}

\noindent {\it Convection and Mixing}--{Time-dependent convection (TDC; \citealt{2023ApJS..265...15J}), using a local implementation of the \cite{1986A&A...160..116K} model, was employed to describe convective regions.} We vary the mixing length coefficient ($\ell = $ mixing length, and $H_p =$ pressure scale height) $\alpha_{\rm MLT}\equiv\ell/H_p$ as $1.5, 2.0$ ({\it canonical}) and $2.5$ (see \S\ref{sec:MLT}). Semi-convection is included via the Langer-El Eid-Fricke criterion \citep{1985A&A...145..179L} with efficiency parameter $\alpha_{\rm semi} = 0.1$. We investigated the influence of varying $\alpha_{\rm semi} = 1$ and $10$ in \S\ref{sec:semiconv}. We note that thermohaline mixing ($\theta$) has negligible effects on the global evolutionary properties of Pop~III stars, with variations in $\theta$ from 0 to 1000 producing virtually identical H-R tracks, evolutionary timescales, and stellar structure parameters, indicating that this mixing process is not a dominant factor in the evolution of low-mass primordial stars. In the {\it canonical} models discussed below, $\theta \equiv 0$. 

\noindent {\it Element Diffusion and Overshooting}---Microscopic diffusion is included using the \cite{1969fecg.book.....B} formalism (see also \citealt{2018ApJS..234...34P}), accounting for gravitational settling and thermal diffusion of elements. To assess the influence of convective boundary mixing on the evolutionary outcomes of Pop~III stars, we computed models employing step overshooting \citep{2000A&A...360..952H} with two distinct efficiency combinations. In the first set, we adopt $\alpha_{\rm ov } = 0.15$ for both hydrogen- and helium-burning regions, encompassing cores and shells, with $\alpha_{\rm ov, 0} = 0.01$ plus exponential $f_{\rm ov} = 0.003$ and $f_{\rm ov,0} = 0.001$ on top of any other convective core. For comparison, we also computed a {\it control} model without any overshooting and a {\it canonical} model---presented in the next section---with moderate step overshooting $\alpha_{\rm ov~, H} = 0.11$ at the boundary of the convective hydrogen-burning core, when present, $\alpha_{\rm ov~, He} = 0.06$ the boundary of the convective helium-burning core, both with $\alpha_{\rm ov,0} = 0.1$, plus exponential $f_{\rm ov} = 0.005$ and $f_{\rm ov,0} = 0.001$. 

\noindent {\it Nuclear Reaction Network}---We used a 30-isotope network including light elements up to $^{56}{\rm Fe}$ (dubbed {\tt mesa$\_$30.net}). Iron in the low-mass reaction network does not actively participate in nuclear reactions; rather, it is included solely to ensure a consistent specification of the initial composition and mass conservation. In the present cases, the initial iron abundance is negligible, with its inclusion therefore relevant only for microphysical processes, such as opacity, the equation of state and diffusion. Reaction rates were drawn from JINA REACLIB \citep{2010ApJS..189..240C}, NACRE I and II \citep{1999NuPhA.656....3A, 2017RvMP...89c5007D, 2022ApJ...935...21C}, with weak rates from \cite{1994ADNDT..56..231O, 2000NuPhA.673..481L}. Electron screening follows \cite{2007PhRvD..76b5028C}, and thermal neutrino losses were treated following \cite{1989ApJ...339..354I, 1992ApJ...395..622I, 1996ApJS..102..411I, 1996ApJ...470.1015I}.

\section{Evolution of Low-Mass Pop~III Stars}\label{sec:evolutionary_tracks}

\subsection{General Characteristics}

\begin{figure}[t]
    \centering
    \includegraphics[width = \linewidth]{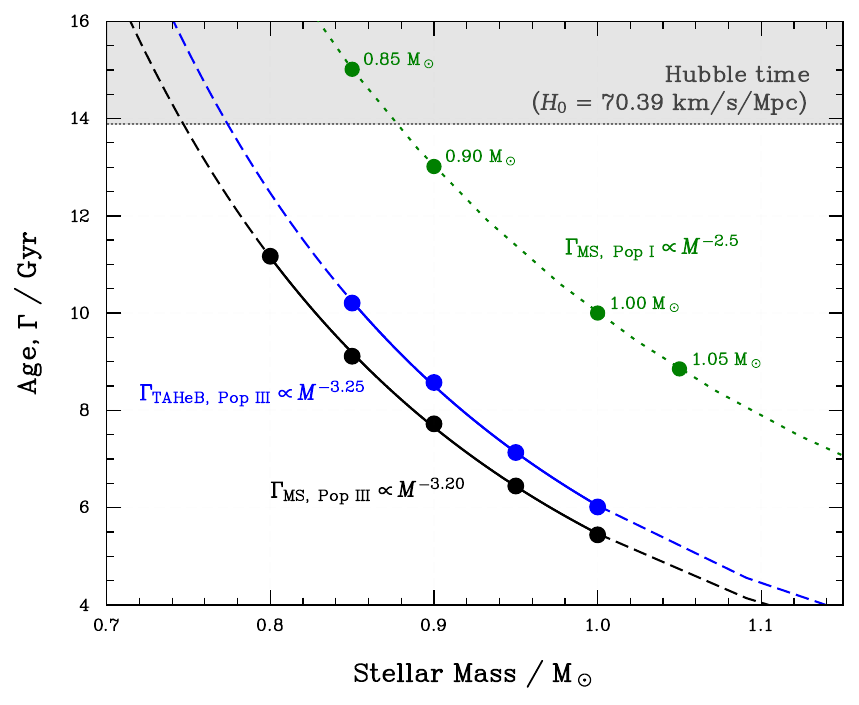}
    \caption{Main sequence lifetimes ($\Gamma_{\rm MS}$) of low-mass Pop~III models follow a simple power-law $\propto M^{-3.2}$, steeper than the canonical Population I relation ($\propto M^{-2.5}$; green dashed line). Similarly, the terminal age of central helium burning (TAHeB; defined here as the point when $X_{^{4}{\rm He}} < 0.01$) obeys a comparable scaling, $\propto M^{-3.25}$ (blue solid line). The horizontal dashed line marks the Hubble time for $H_0 = 70.39~K$m s$^{-1}$ Mpc$^{-1}$ following \cite{2025ApJ...985..203F}.}
    \label{fig:MSlifetimes}
\end{figure}

Accurate main-sequence lifetimes are critical for constraining early-Universe chemical feedback, identifying long-lived Pop~III survivors, and probing near-field cosmology. Low-mass Pop~III stars follow a steeper power law, $\Gamma_{\rm MS, ~Pop~III} \propto \left(M/M_\odot\right)^{-3.2}$ compared to Pop~I scaling $\Gamma_{\rm MS} \propto \left(M/M_\odot\right)^{-2.5}$ (Figure~\ref{fig:MSlifetimes}). At MSTO, ${Z = 10^{-3}}$ models have helium cores of $\sim 0.03~M_\odot${, defined as the outermost mass coordinate where the hydrogen fraction $X_c$ drops below 0.1}, whereas $Z = 0$ models reach $\sim 0.1~M_\odot$, reflecting slower hydrogen processing. {In metal-free stars, the absence of CNO nuclei at formation precludes energy generation via the CNO cycle, so hydrogen burning proceeds initially through the pp chain ($\epsilon_{\rm pp} \propto T^4$) rather than the CNO cycle ($\epsilon_{\rm CNO} \propto T^{17}$; \citealt{2004sipp.book.....H}).} {Such weaker temperature sensitivity reduces central temperatures, producing an extended radiative core and minimal convective mixing, which slows hydrogen consumption and extends the main-sequence lifetimes of low-mass Pop III stars, consequently delaying both chemical enrichment and radiative feedback relative to metal-enriched counterparts.}

A complementary diagnostic is the total timescale for a $Z = 0$ star to evolve into a white dwarf cooled to $L \simeq 1~L_\odot$, which naturally declines sharply with mass, from $12.36$ Gyr for $0.8~M_\odot$ to $6.02$ Gyr for $1.0~M_\odot$, with intermediate values of $10.23$, $8.59$, and $7.14$ Gyr for $0.85$, $0.90$, and $0.95~M_\odot$, respectively, implying that only metal-free models with $M \lesssim 0.8~M_\odot$ may remain unevolved today, consistent with dynamical simulations \citep{2013MNRAS.435.3283M, 2016MNRAS.462.1307S, 2020ApJ...901...16D, 2022ApJ...925...28L}, whilst more massive progenitors have already produced white dwarfs. {We notice that gravitational settling in our models increases the central helium fraction and mean molecular weight, raising the core temperature and accelerating pp-chain burning. Additionally, small variations in the treatment of convective boundaries likewise affect the efficiency with which hydrogen is mixed toward the core, subtly modifying the rate of nuclear exhaustion. Together, these effects yield slightly shorter evolutionary timescales relative to \cite{2023MNRAS.525.4700L}---see Table \ref{tab:IIIWDs}---, although the resulting WD ages remain consistent at the $\sim7-12\%$ level.}

\begin{figure*}[t]
    \centering
    \includegraphics[width = \linewidth]{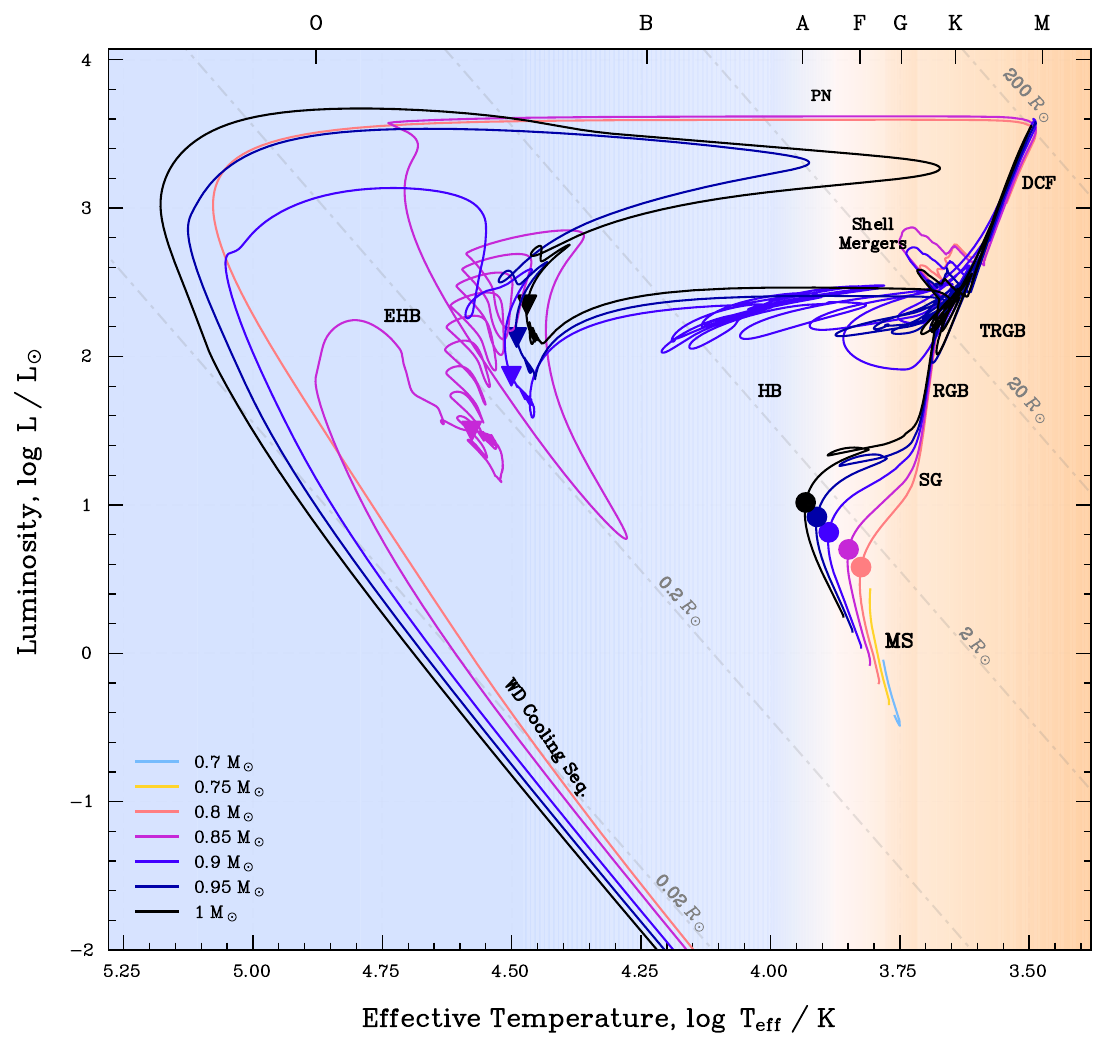}
    \caption{Evolutionary tracks for Pop~III stellar models with ZAMS masses ranging from $0.7$ to $1.0~M_\odot$. Models with $0.7-0.75~M_\odot$ do not leave the MS within the Hubble time. Dashed lines indicate constant stellar radii (0.02, 0.2, 2, 20, 200 $R_\odot$, from left to right). Background colours indicate spectral types. Filled circles mark the point at which the central hydrogen mass fraction drops below 0.01 ($X_c < 0.01$), whilst inverted triangles indicate when the central helium mass fraction falls below 0.01 ($Y_c < 0.01$) during helium burning. The main evolutionary phases---Main Sequence (MS), Subgiant (SG), Red Giant Branch (RGB), Tip of the RGB (TRGB), {Double Core Flash (DCF)}, Horizontal Branch (HB), Planetary Nebula (PN), Extended Horizontal Branch (EHB), and White Dwarf (WD) cooling sequence---are indicated.}
    \label{fig:HR}
\end{figure*}

Figure~\ref{fig:HR} displays the H-R diagram for the set of low-mass Pop~III stellar models analysed in this study, covering masses from $0.7$ to $1.0~M_\odot$. The evolutionary tracks extend from the Zero Age Main Sequence (ZAMS) to the WD phase for models with initial masses above $M_{\rm ZAMS} \geq 0.8~M_\odot$. We highlight the onset of core helium ignition, shell-burning episodes, and the eventual cooling and contraction leading to the WD endpoint.

\subsection{Metallicity-Driven Structural Evolution}\label{sec:comparison_highZ}

\begin{figure}[t]
    \centering
    \includegraphics[width = \linewidth]{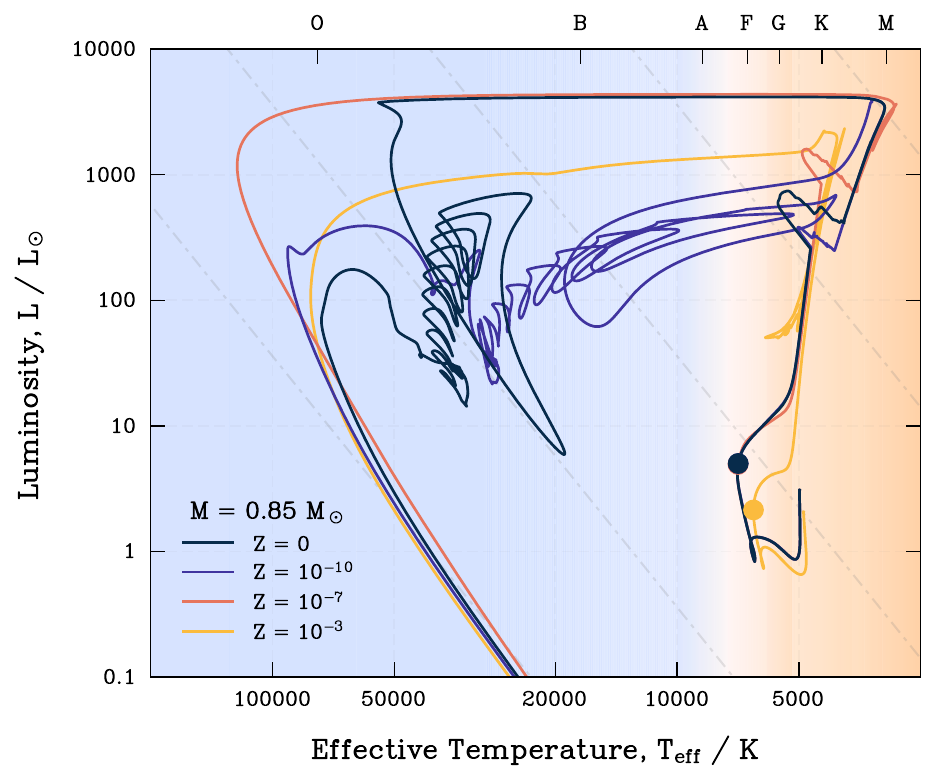}
    \caption{Stellar evolution tracks for $0.85~M_\odot$ models across different metallicities from Pop~III ($Z = 0$) to metal-poor Pop~II stars ($Z = 10^{-3}$). {Coloured dots indicate the MSTO}.}
    \label{fig:multi_metal}
\end{figure}

\begin{figure*}[t]
    \centering 
    \includegraphics[width = \linewidth]{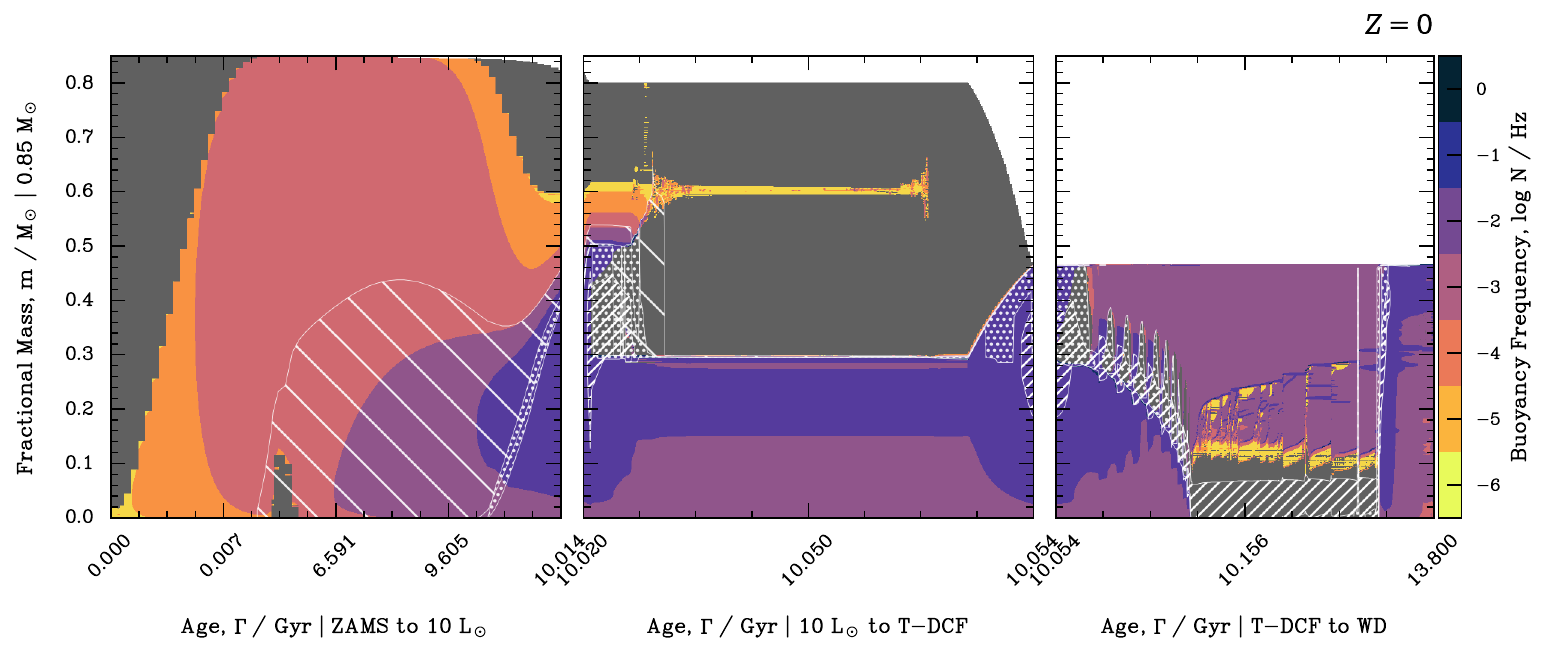}
    \includegraphics[width = \linewidth]{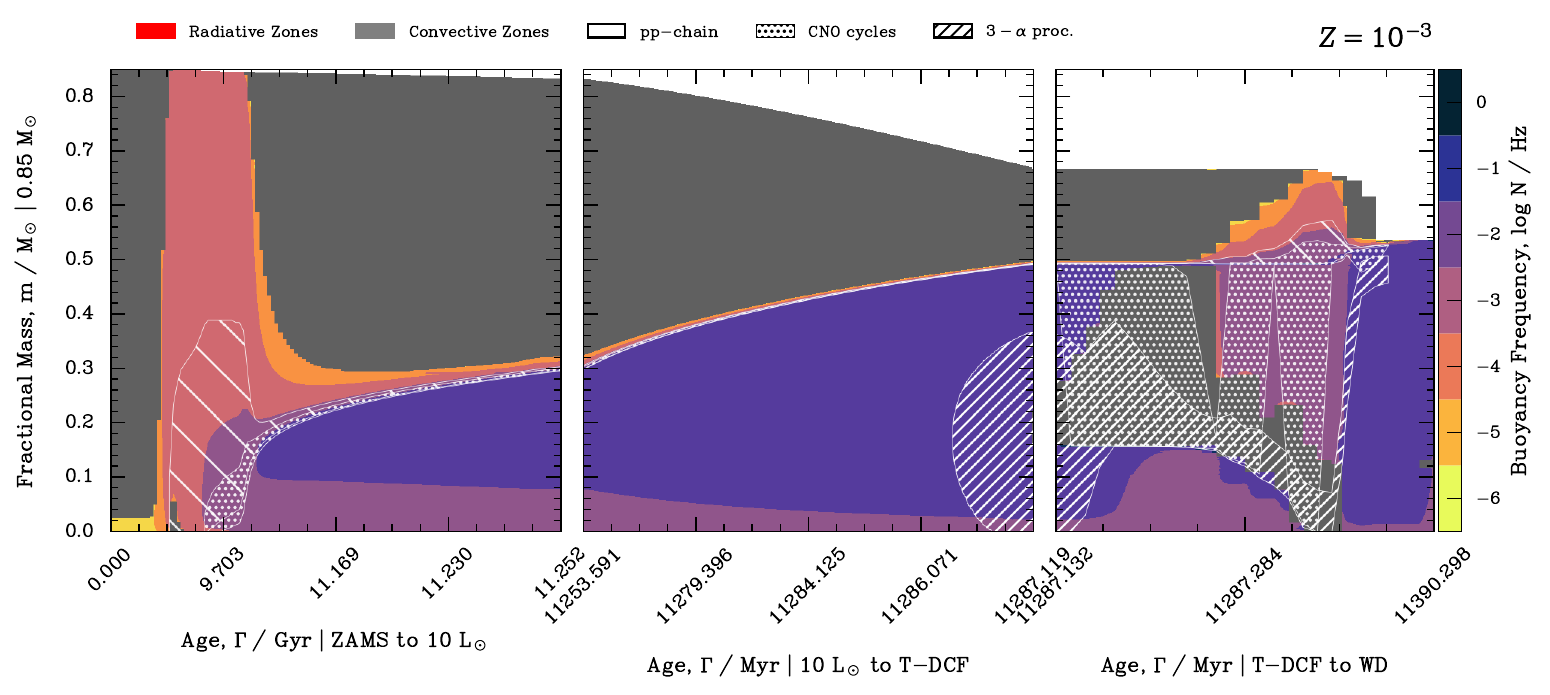}
    \caption{Kippenhahn diagrams showing the internal structural evolution of $0.85~M_\odot$ stellar models as a function of age for $Z = 0$ (first panel) and $Z = 10^{-3}$ (second panel). Each set is presented across three evolutionary intervals: from the zero-age main sequence (ZAMS) to the point where the star attains $10~L_\odot$ (left), from this stage to the tip of the DCF (centre), and from there through to the WD phase (right). Radiative zones are depicted using a colour scale, whilst convective regions (where the Brunt-V{\"a}is{\"a}l{\"a} frequency satisfies $N^2 < 0$) are shown in grey. Nuclear burning regions are marked using distinct hatching: leftward diagonal lines for the pp-chain, dots for the CNO cycle, and rightward diagonal lines for the $3\alpha$ process.}
    \label{fig:kippmetals}
\end{figure*}

The introduction of even trace metallicities in low-mass stellar models produces coupled structural and evolutionary changes, altering energy generation, mixing, and envelope dynamics. These arise from feedback between nuclear burning, opacity-driven transport, and degeneracy physics: metallicity raises opacity, which steepens the radiative gradient, modifying convection and hence the mixing that reshapes the burning environment. Such effects are evident in the H–R and Kippenhahn diagrams for our $0.85~M_\odot$ tracks presented in Figures~\ref{fig:multi_metal} and \ref{fig:kippmetals}. 

All models begin core hydrogen burning through the pp-chain ($\epsilon_{\rm pp} \propto T^4$). For $Z \gtrsim 10^{-10}$, trace CNO nuclei enable the CNO cycle at later stages, whose steep temperature sensitivity drives higher core luminosities and modest convective cores \citep[e.g.][]{1990ApJ...349..580F}. Such convective mixing replenishes central hydrogen and smooths gradients, extending the main-sequence lifetime. In contrast, metal-free models sustain lower central temperatures and shallower gradients, inhibiting convective core formation. Their helium cores thus emerge as compact, chemically stratified regions beneath predominantly radiative interiors. 

As the models evolve off the main sequence, metallicity-dependent opacities drive distinct RGB morphologies \citep{2005essp.book.....S, 2021A&A...646L...6H}. At these regimes, the Rosseland opacity is dominated by H$^-$ bound-free absorption, which scales with free electron density supplied by metals; higher $Z$ therefore steepens envelope opacity. {Although a shallow outer convective layer persists beyond the RGB in all models due to partial H-ionisation, its radial extent is strongly metallicity dependent and remains insufficient in Pop~III stars to connect with the burning shells.} Consequently, metal-rich models undergo strong RGB expansion, whilst the Pop~III track remains blue and compact, {with envelope convection that is present but too shallow to induce early dredge-up}, and delaying surface abundance changes and suppresses early spectroscopic evidence of internal burning. The $Z = 10^{-7}$ track, on the other side, exhibits intermediate deepening and milder expansion. Metallicity also dictates helium ignition, whereas at $Z = 10^{-3}$, helium ignites at $\rho_c \sim 10^{5.2}~{\rm g~cm^{-3}}$ under moderate degeneracy, producing a single, mild flash with limited convective penetration. Higher envelope opacity insulates the core, reducing neutrino losses and promoting stable helium burning (Figure~\ref{fig:kippmetals}). Conversely, in $Z = 0$ models, enhanced neutrino cooling shifts $T_{\rm max}$ off-centre, and helium ignites at $\rho_c \gtrsim 10^6~{\rm g~cm^{-3}}$ under near-complete degeneracy. Expansion is decoupled from temperature, so once the $3\alpha$ rate rises, runaway burning ensues. This drives a violent, off-centre flash with deep, episodic convective zones, reinforced by further plasma and photo-neutrino cooling \citep{1996ApJ...470.1015I}. The $Z = 10^{-7}$ model again displays intermediate ignition densities and flash strengths. 

Post-flash behaviour diverges further, with the Pop III model developing a hotter, more compact helium-burning shell, with higher peak temperatures and steeper entropy gradients. Reduced steady-state H-shell luminosity allows greater helium accumulation, strengthening thermal pulses, and recurrent, high-amplitude shell instabilities appear during the pre-core-helium-burning phase (Figure~\ref{fig:kippmetals}). {We call attention, however, that the luminosity excursions observed before the onset of stable core helium burning are not classical Asymptotic Giant Branch thermal pulses, but correspond to a sequence of inward-propagating helium-burning instabilities, as indicated by the progressively decreasing mass coordinate of the ignition region in Figure \ref{fig:kippmetals}. In the Pop~III case, strong degeneracy in the helium-rich layers and steep entropy gradients favour discrete flash-like events, which drive the looping behaviour observed in the HR diagram at the TRGB and before the HB/EHB phases.} In contrast, $Z = 10^{-3}$ models show weaker or absent pulses owing to reduced helium-shell degeneracy and associated radiative energy transport, whilst $Z = 10^{-7}$ stars occupy a transitional regime. {For the $Z = 10^{-3}$ models, the smoother behaviour reflects reduced helium-shell degeneracy and shallower entropy gradients associated with higher opacity, allowing the helium-burning front to migrate inward quasi-continuously without triggering discrete flash events; $Z = 10^{-7}$ models exhibit intermediate behaviour between these two regimes.} {Despite originating as metal-free, our Pop~III stellar models undergo self-enrichment during late stages, with thermal pulses driving interior shell mergers and deep convective dredge-up, carrying $^{12}$C from $3\alpha$ burning and $^{14}$N from CN-cycles to the surface (see Appendix~\ref{app:evolution}). By contrast, higher-metallicity stars experience weaker late dredge-up, having already undergone the first dredge-up on the RGB. Table~\ref{tab:PopIIICNO} summarises the resulting total CNO mass fractions ${\rm X}_{\rm CNO}$ alongside the surface abundance after flash reported by \cite{2023MNRAS.525.4700L}.} The resulting Pop~III WD is, therefore, distinct, forming a hotter, denser, and more compact core, with a thinner H envelope, reduced residual burning, and delayed crystallisation \citep{1968ApJ...151..227V, 2019MNRAS.482.5222T}. Its cooling sequence is correspondingly faster, with potentially distinctive asteroseismic signatures.

\subsection{Impact of Mass-Loss Efficiency}\label{sec:mass-loss}

Mass loss remains one of the least constrained parameters in low-mass stellar evolution, especially at the lowest metallicities. {Unlike metal-rich stars, Pop III models lack initial surface metals to drive radiative winds, leaving RGB envelope stripping uncertain (cf. \citealt{2023MNRAS.525.4700L}, who adopted $\eta_{\rm R}=0.3/0.378$ for the Reimers prescription and applied an additional dust-driven mass-loss law calibrated to the \citealt{2018A&A...609A.114G} measurements of pulsating giant stars).} Nonetheless, dredge-up and shell mergers may trigger outflows, as discussed in \S\ref{sec:085} (see also \S\ref{sec:08}, \ref{sec:09}, \ref{sec:095-1}). Current prescriptions rely on empirical Pop~I/II relations (e.g., \citealt{1977A&A....61..217R, 1990A&A...231..134N, 1995A&A...299..755B, 1988A&AS...72..259D, 2005A&A...438..273V}), but their validity at $Z\longrightarrow0$ is uncertain. Recent work suggests that for metallicities ${\rm [M/H]} < -1$, mass-loss rates may fall by orders of magnitude, with pulsation- and shock-driven outflows highly sensitive to envelope opacity, chromospheric dissipation, and acoustic leakage (e.g., \citealt{2025ApJ...988..179L}, their Fig.~7). Observational constraints for EMP giants remain inconclusive, and it is unknown whether Pop~III stars expel their envelopes at all. Nevertheless, we evolved $0.85~M_\odot$ Pop~III models under Reimers prescriptions with $\eta_{\rm R} = 0.1$, $0.5$, $0.7$, and $0.9$, holding the Bl{\"o}cker term fixed on the DCF. Diagnostics of core growth, envelope evolution, and surface enrichment were tracked from pre-MS to WD cooling. 

\begin{figure}[t]
    \centering
    \includegraphics[width = \linewidth]{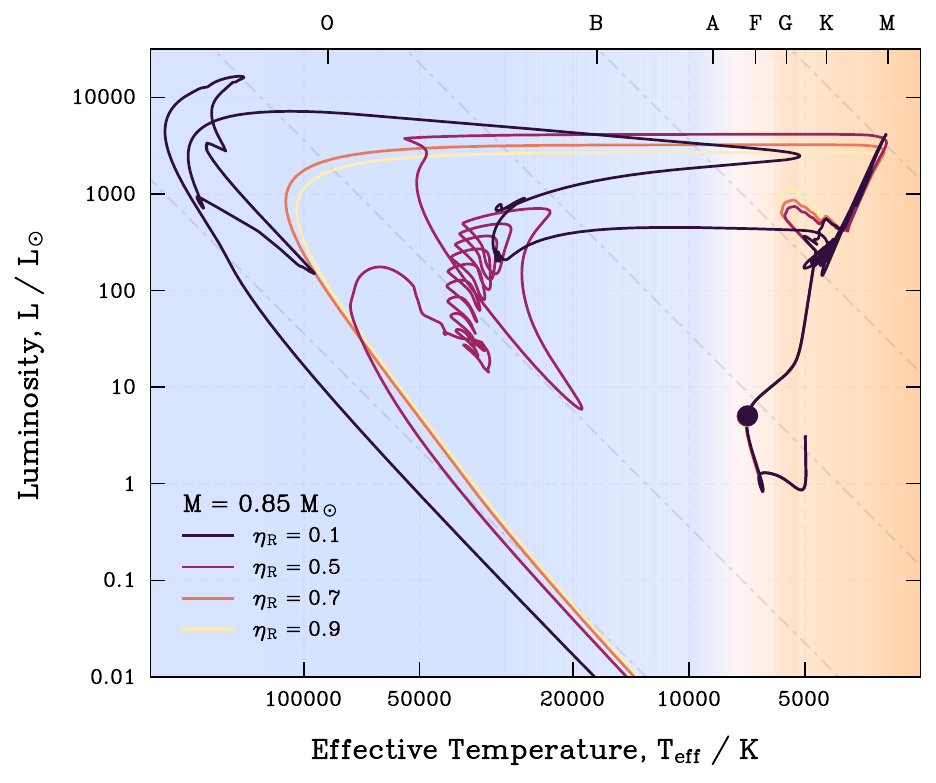}
    \caption{H–R diagram for $0.85~M_\odot$ Pop~III models with varied Reimers mass-loss efficiencies $\eta_{\rm R} = 0.1$, 0.5, 0.7, 0.9, evolved from pre-MS to WD cooling. For $\eta_{\rm R} \leq 0.1$, stars retain most of their envelopes, following extended RGB–DCF tracks and ending as relatively cool post-DCF objects ($T_{\rm eff} \lesssim 30~000~K$). For $\eta_{\rm R} \geq 0.5$, pre-helium-ignition envelope stripping produces EHB-like morphologies ($T_{\rm eff} \gtrsim 40~000~K$) with truncated luminosity excursions. The bifurcation is sharp and highly $\eta_{\rm R}-$dependent, whilst pre-dredge-up metallicity remains negligible.}
    \label{fig:MassLossCMD}
\end{figure}

Figure~\ref{fig:MassLossCMD} shows the resulting H-R diagrams. With $\eta_{\rm R} = 0.1$, peak RGB rates remain low ($\dot{M} \sim 4.2 \times 10^{-7}~M_\odot~{\rm yr}^{-1}$), yielding only $0.12~M_\odot$ ejected mass. The hydrogen-rich envelope survives the helium flash, producing red/yellow HB morphologies akin to $0.95-1~M_\odot$ Pop~III stars (\S\ref{sec:095-1}) and a $0.73~M_\odot$ WD. In contrast, $\eta_{\rm R} = 0.7-0.9$ cases lose $0.41-0.43~M_\odot$ by the TRGB ($\dot{M} \sim 3.3-3.4 \times 10^{-7}~M_\odot~{\rm yr}^{-1}$), reducing the envelope to $M_{\rm env} \sim 10^{-4}~M_\odot$. These stars truncate the T-DCF, ignite helium under compact conditions, evolve along EHB-like tracks, and reach $T_{\rm eff} \gtrsim 2\times10^4~K$. Their remnants are pure He-core WDs of $M_{\rm core, He} \approx 0.43-0.44~M_\odot$, with envelopes well below the canonical $10^{-2}~M_\odot$ DA threshold.  

\begin{figure}[t]
    \centering
    \includegraphics[width = \linewidth]{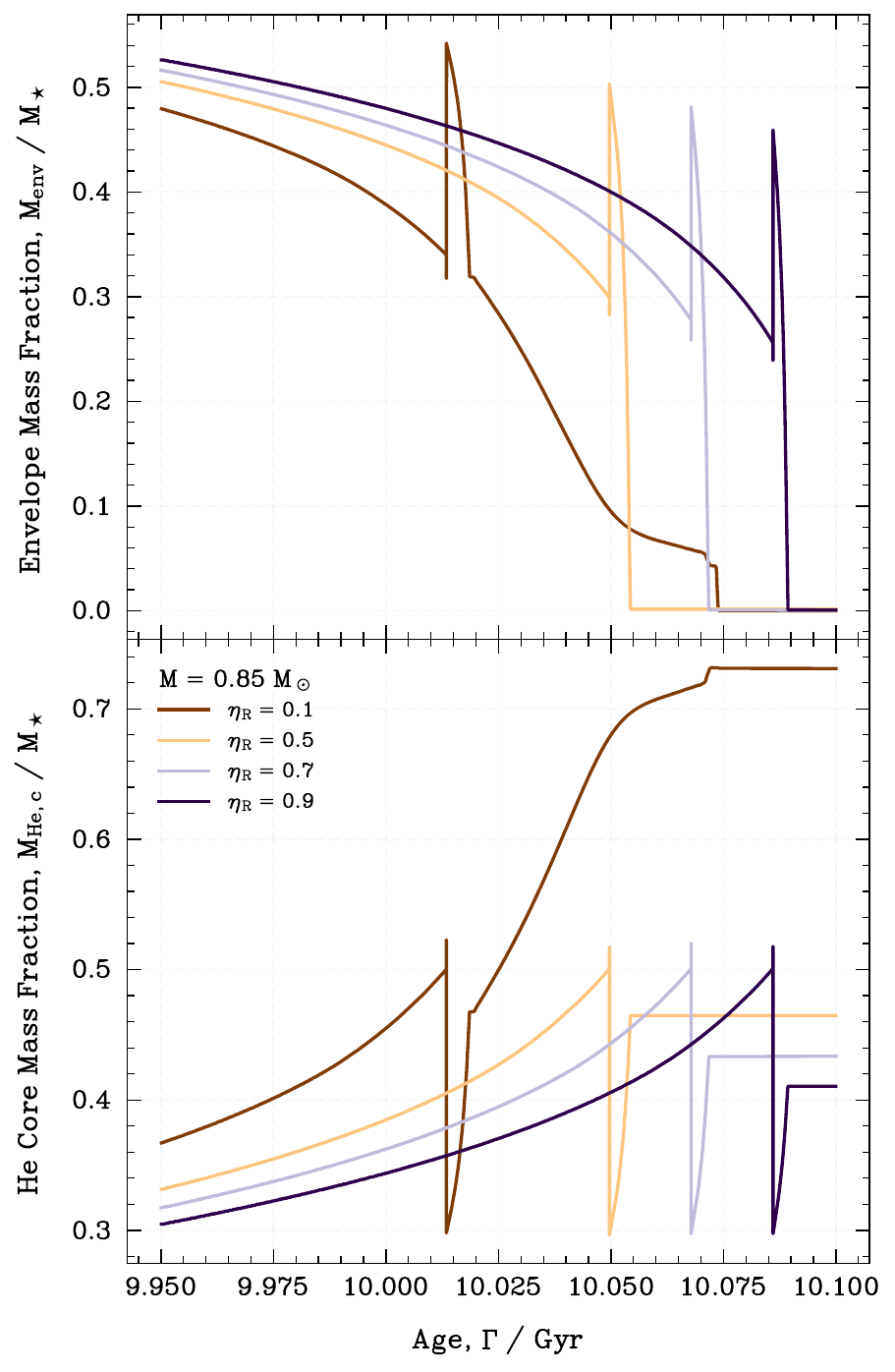}
    \caption{Time evolution of the hydrogen envelope fraction ($M_{\rm env}/M_\star$; upper) and helium core fraction ($M_{\rm He,c}/M_\star$; lower) during the TRGB and early DCF for $0.85~M_\odot$ Pop~III models with $\eta_{\rm R} = 0.1$, 0.5, 0.7, 0.9. Helium core growth is largely insensitive to $\eta_{\rm R}$, whereas the envelope declines sharply for $\eta_{\rm R} \gtrsim 0.5$, reaching $< 10^{-3}~M_\odot$. This rapid depletion marks the onset of stripped-envelope evolution and bifurcation in the final WD structure and thermal history.}
    \label{fig:MassLossEnvelopeCoreMass}
\end{figure}

Figure~\ref{fig:MassLossEnvelopeCoreMass} highlights such divergent behaviour. The He-core mass---defined where the hydrogen fraction drops below 0.01---grows smoothly with $\eta_{\rm R}$, differing by $<2\%$ across cases. Envelope masses, however, respond non-linearly, with the $\eta_{\rm R} = 0.1$ track retaining most of its hydrogen, whereas $\eta_{\rm R} = 0.7-0.9$ tracks deplete $M_{\rm env} < 10^{-3}~M_\odot$ by $\Gamma \sim 10$ Gyr. This defines a critical transition near $\eta_{\rm R} \sim 0.4-0.6$ between classical post-DCF and stripped-envelope outcomes, and such sensitivity may directly impact UV flux, dredge-up efficiency, and the observability of Pop~III pre-WDs and WDs in photometric and asteroseismic channels.  

\subsection{Impact of Convective Transport Efficiency}\label{sec:MLT}

Convection in stellar models is commonly described via mixing-length theory (MLT; \citealt{1958ZA.....46..108B}), where convective efficiency is parametrised by $\alpha_{\rm MLT} = \ell/H_P$, the ratio of mixing length $\ell$ to local pressure scale height $H_P$. Solar-calibrated values lie in $\alpha_{\rm MLT} \approx 1.6–2.1$ (e.g., \citealt{2015A&A...573A..89M, 2019MNRAS.489.4712S, 2022RNAAS...6...77C}), but their extrapolation to metal-free populations is unconstrained and computationally sensitive \citep{2020MNRAS.492.4986Y, 2022ApJ...939...28C}. In low-mass Pop~III models, compact, radiatively transparent envelopes make the local temperature gradient, convective boundaries, and post-MS structural response strongly sensitive to $\alpha_{\rm MLT}$ \citep{1999ApJ...526..991A, 2004MmSAI..75..300A, 2009IAUS..259..233S}.  

\begin{figure}[t]
    \centering
    \includegraphics[width=\linewidth]{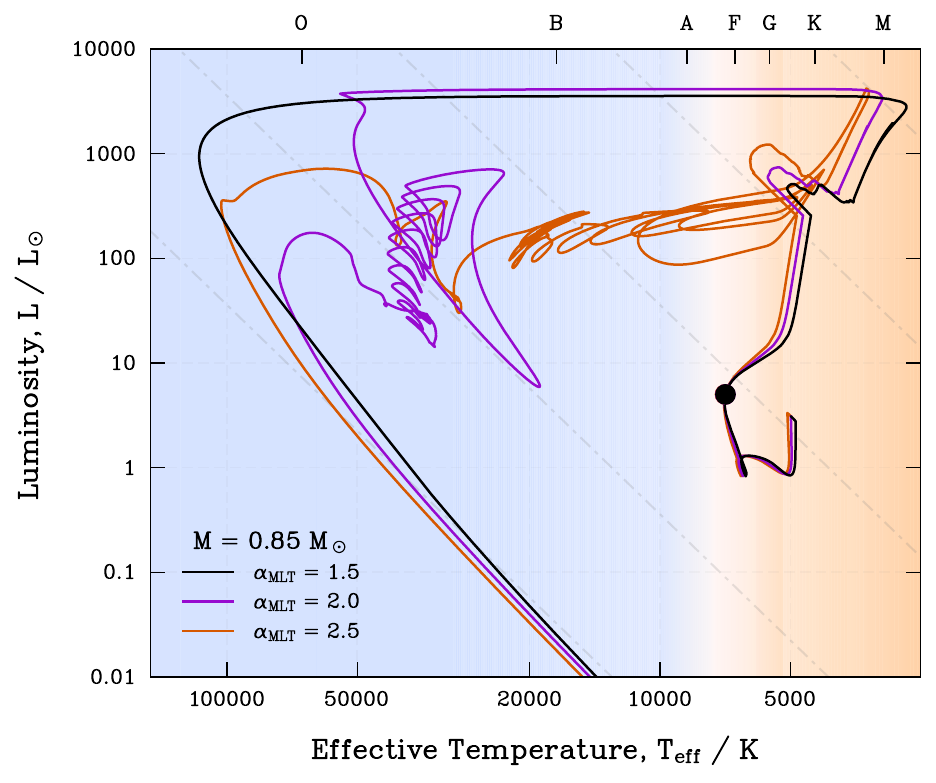}
    \caption{H–R diagram for $0.85~M_\odot$ Pop~III models with varying $\alpha_{\rm MLT}$. Higher $\alpha$ enhances convective transport, producing slightly hotter RGB structures. Post-RGB, $\alpha_{\rm MLT} = 2.5$ models evolve into hotter, more extended EHB tracks, whilst $\alpha_{\rm MLT} = 1.5$ models remain redder. Post-DCF/pre-WD tracks show corresponding temperature shifts.}
    \label{fig:CMD_MLT}
\end{figure}

Figure~\ref{fig:CMD_MLT} illustrates the H–R response of $0.85~M_\odot$ Pop~III models with $\alpha_{\rm MLT} = 1.5, 2.0, 2.5$. Higher values of $\alpha_{\rm MLT}$ enhance convective transport, reducing the super-adiabatic gradient $\nabla - \nabla_{\rm ad}$ and producing slightly higher $T_{\rm eff}$ during the RGB. Post-RGB, differences amplify: $\alpha_{\rm MLT} = 2.5$ models undergo pronounced blueward excursions after helium ignition, forming hotter, more extended EHB-like morphologies, whereas $\alpha_{\rm MLT} = 1.5$ tracks remain cooler and redder. These structural differences arise from thinner convective envelopes and reduced entropy gradients in high$-\alpha$ models, yielding compact post-flash configurations and higher surface temperatures. Post-DCF and pre-WD phases show corresponding shifts in temperature, affecting UV output and potentially the pulsation spectra of WD remnants \citep{1993ApJ...419..596D, 2001PASP..113.1162M}.  

\begin{figure}[t]
    \centering
    \includegraphics[width=\linewidth]{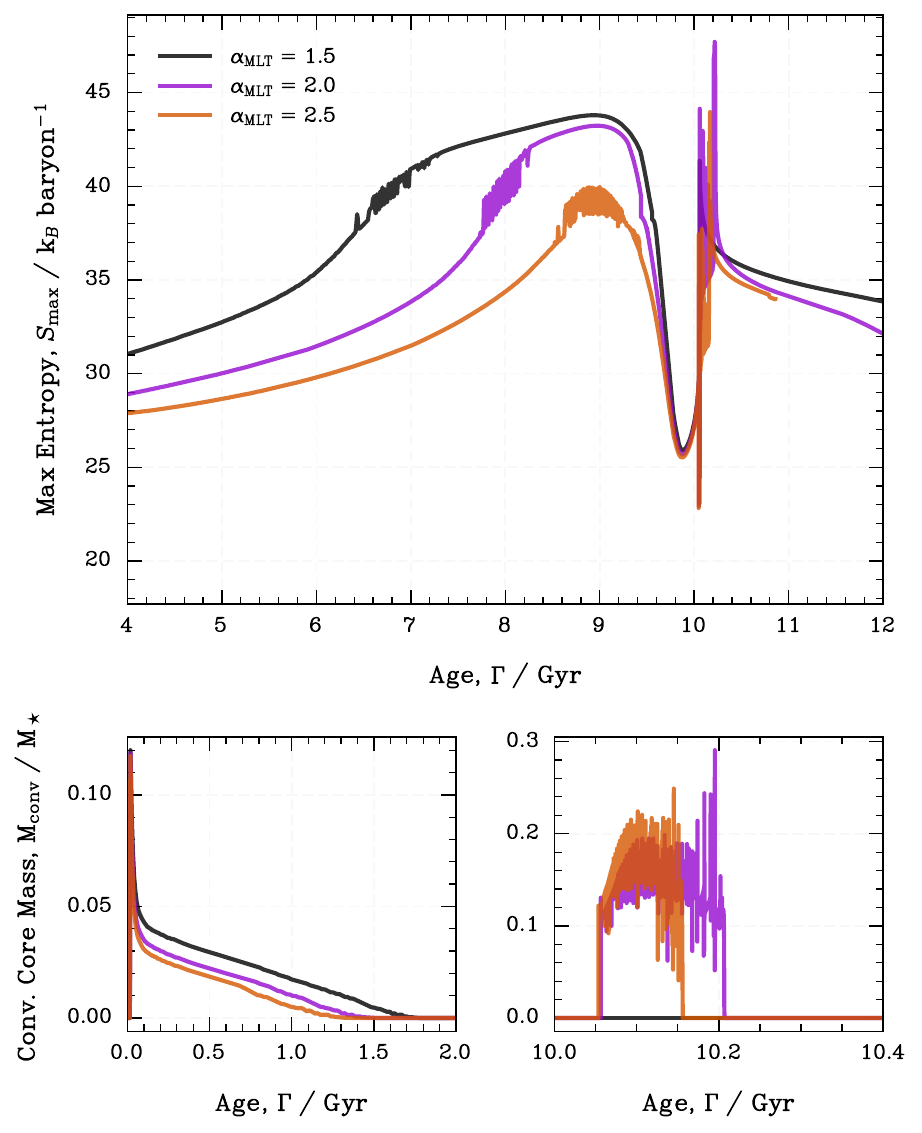}
    \caption{{\it First panel}: Maximum entropy with varying $\alpha_{\rm MLT}$ for a $0.85~M_\odot$ Pop~III stellar model. The profile shows three distinct phases: (i) a gradual increase during MS as the core becomes more degenerate, (ii) a steep rise during RGB ascent as the envelope expands and becomes more convective, and (iii) a sharp drop followed by recovery during the He-flash when core neutrino cooling temporarily dominates. Higher $\alpha_{\rm MLT}$ leads to more efficient convective energy transport and systematically lower entropy values. {\it Second panel}: Convective core mass evolution during early MS ($0–2$~Gyr; left) shows minimal sensitivity to $\alpha$, whilst near the RGB tip ($10.0-10.4$ Gyr; right) subtle variations emerge as envelope convection modifies interior flux and convective boundaries.}
    \label{fig:varyMLT}
\end{figure}

Panel diagnostics in Figure~\ref{fig:varyMLT} reveal how $\alpha_{\rm MLT}$ variations propagate through the interior structure. Convective core masses during early MS hydrogen burning ($0$–2~Gyr) remain largely insensitive to envelope MLT, being set by nuclear energy generation $\epsilon_{\rm nuc} \propto \rho T^\kappa$ and opacity via the Schwarzschild criterion. Approaching helium ignition ($10.0-10.4$ Gyr), envelope convection subtly modifies interior flux and pressure waves, producing small variations ($\lesssim 0.02$ in $M_{\rm conv}/M_\star$).  These structural differences may also impact asteroseismic observables through changes in the acoustic cavity, since large frequency separations scale as a function of mass and radius of a star, whilst small separations are proportional to the sound-speed propagation within their interiors, probing core-envelope coupling via $c_s^2 = \Gamma_1 P/\rho$. Shifts in convective-radiative boundaries affect mode damping and frequency ratios sensitive to envelope stratification \citep{2012A&A...540L...7B, 2013ApJ...769..141S}, offering potential diagnostics of $\alpha_{\rm MLT}$ in Pop~III stars and tests of primordial stellar evolution predictions (Paper II).  

Variations in the maximum entropy evolution also reflect fundamental thermodynamic changes in stellar structure. Models with lower $\alpha_{\rm MLT} = 1.5$ maintain steeper entropy gradients by the He-flash, with these differences altering the acoustic cavity via: (1) direct modification of $c_s$ through the entropy gradient term, and (2) structural changes in $P$ and $\rho$ from convective efficiency variations since pressure-mode frequencies scale directly with $(c_s^2)^{-1}$, so higher entropy in low-$\alpha$ models produces longer acoustic path lengths and lower frequencies. Gravity-modes, on the other side, are set by the Brunt–V{\"a}is{\"a}l{\"a} frequency, so steeper entropy gradients in low-$\alpha$ models enhance buoyancy and modify g-mode spectra. These effects peak post-He-flash, implying that asteroseismic measurements of EHB stars could further constrain convective efficiency in Pop~III stars.

\subsection{Impact of Semi- and Convective Boundary Mixing}\label{sec:semiconv}

Semi-convection is a mixing process occurring in regions where the Schwarzschild criterion predicts instability ($\nabla_{\rm rad} > \nabla_{\rm ad}$), but the stabilising composition gradient renders the layer Ledoux-stable ($\nabla_{\rm rad} < \nabla_{\rm ad} + (\varphi/\delta)\nabla_\mu$; \citealt{1947ApJ...105..305L, 1958ApJ...128..348S, 1966PASJ...18..374K}). Its efficiency is parametrised by $\alpha_{\rm semi}$ \citep{1983A&A...126..207L}, which controls the rate at which composition gradients are smoothed in semi-convective zones. Varying $\alpha_{\rm semi}$ primarily affects convective core growth, adjacent layer structure, and surface CNO abundances, with stronger effects in higher-mass stars \citep{1969MNRAS.144..231T, 1993SSRv...66..409M, 2002RvMP...74.1015W, 2013A&A...552A..76S, 2021A&A...655A..29J}.

We explored the impact of semi-convective efficiency on a $0.85~M_\odot$ Pop~III star by adopting $\alpha_{\rm semi} = 0.1$, 1, and 10. The MS lifetime ($\Gamma_{\rm MS} \approx 9.11~{\rm Gyr}$) and morphology remain essentially invariant. Post-MS evolution, however, exhibits measurable differences in internal structure and surface composition. At the MSTO, the helium core mass ($M_{\rm He} \sim 0.106~M_\odot$) is nearly identical across models. By the TAHeB, $M_{\rm He}$ decreases modestly with increasing $\alpha_{\rm semi}$ ($0.494~M_\odot$ for 0.1, $0.465~M_\odot$ for 1 and 10), whilst the carbon-oxygen core mass ($M_{\rm CO}$) grows slightly with efficiency ($0.377-0.437~M_\odot$), reflecting enhanced helium mixing into burning regions.

Surface abundances post-TAHeB also reveal distinct behaviour (see Figure~\ref{fig:OV_nucleo}, {middle} panel), with $^{12}$C and $^{14}$N remaining largely unchanged, indicating limited sensitivity to semi-convective transport. $^{16}$O, in contrast, shows a non-monotonic trend: peak surface $^{16}$O occurs at intermediate efficiency ($\alpha_{\rm semi} = 1$), whilst lower and higher efficiencies reduce it. {This reflects the fact that, unlike $^{12}$C and $^{14}$N, which rapidly reach CN-cycle equilibrium, $^{16}$O is destroyed through the kinetically limited ON sub-cycle; its abundance therefore depends on the depth and duration of mixing at ON-cycle temperatures rather than on equilibrium nuclear burning.} {This arises because intermediate mixing allows oxygen-rich material from partially processed layers to be transported to the envelope before prolonged exposure to ON-cycle temperatures, whereas low efficiency limits transport and high efficiency increases the residence time of material in hot burning regions, leading to enhanced $^{16}$O destruction.} Global stellar parameters at TAHeB reflect these compositional differences: higher $\alpha_{\rm semi}$ yields slightly higher luminosities ($\log(L/L_\odot) = -3.83$ to $-2.90$), hotter effective temperatures ($\log(T_{\rm eff}) = 3.72$ to $3.95$), and modestly larger convective core masses during helium burning. All of these changes influence the final WD remnant, particularly the C/O ratio, with consequences for cooling behaviour and perhaps contributions to early Galactic chemical enrichment.

Convective boundary mixing (CBM), typically parametrised via overshooting prescriptions, also represents one of the most significant uncertainties in stellar evolution. CBM influences convective core growth, prolongs the main-sequence lifetime, and shifts stars to higher luminosities and cooler effective temperatures by enhancing fuel mixing and modifying the mean molecular weight \citep{2007ApJ...667..448M}. In one-dimensional stellar evolution models, CBM is commonly implemented either as an exponential overshoot parameter $f_{\rm ov}$ or a step-overshoot parameter $\alpha_{\rm ov}$, {related approximately by $\alpha_{\rm ov}/f_{\rm ov} \approx 20$ for red giants  \citep{2022ApJ...931..116L}.} In regions affected by CBM, the evolution of chemical composition is typically modelled as a diffusive process, with an additional term, $D_{\rm mix}$, capturing the transport of material beyond the formal convective boundary \citep{2023Galax..11...56A}. Hydrodynamic simulations further indicate that convective flows penetrate these boundaries through momentum-driven mixing, with smaller convective cores exhibiting correspondingly reduced overshoot \citep{2021A&A...646A.133H}. In massive stars, variations in CBM significantly alter evolutionary outcomes: extended CBM broadens the main sequence, modifies nucleosynthetic yields (e.g., $^{12}{\rm C}/^{16}{\rm O}$ ratios and weak $s$-process abundances), and affects the convergence of convective boundaries during H- and He-burning \citep{2020MNRAS.496.1967K}. For metal-free Pop~III stars, however, where CNO burning becomes efficient at central temperatures $T_c \geq 2\times10^{7}$~K, CBM governs the transport of nucleosynthetic products to the envelope, influencing evolutionary timescales and surface abundances. Whilst dedicated CBM studies for Pop~III stars are limited, prior investigations comparing Schwarzschild and Ledoux convection criteria demonstrate that convective treatment substantially modifies envelope expansion, dredge-up, and post-He-burning surface composition (e.g., \citealt{2015MNRAS.450.1618L}).  

\begin{figure}[t]
    \centering
    \includegraphics[width = \linewidth]{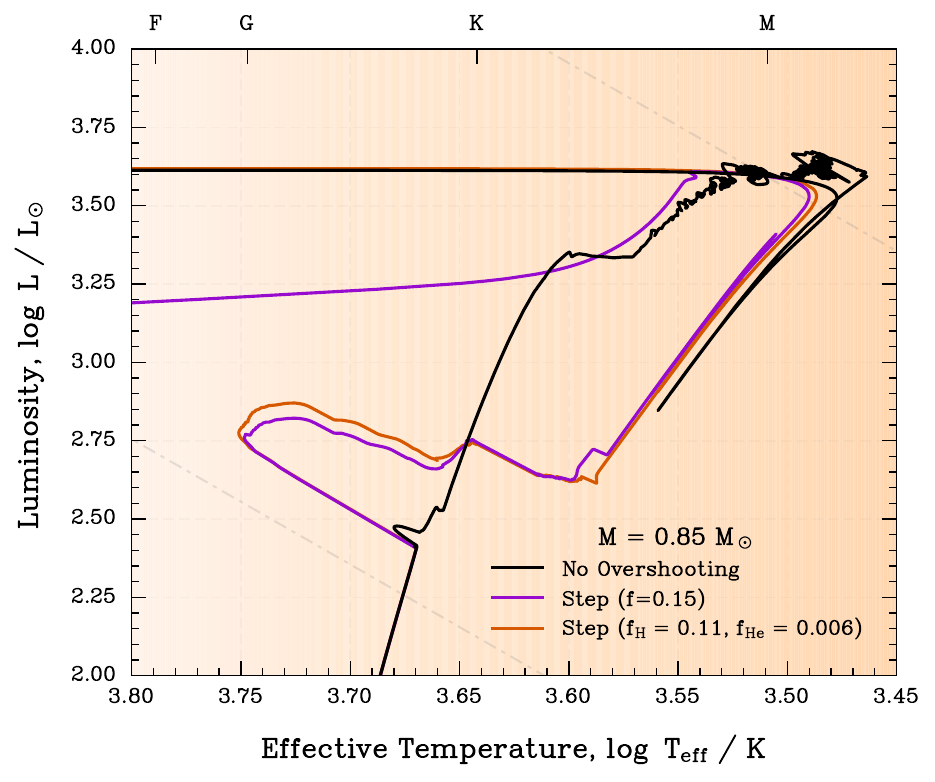}
    \includegraphics[width = \linewidth]{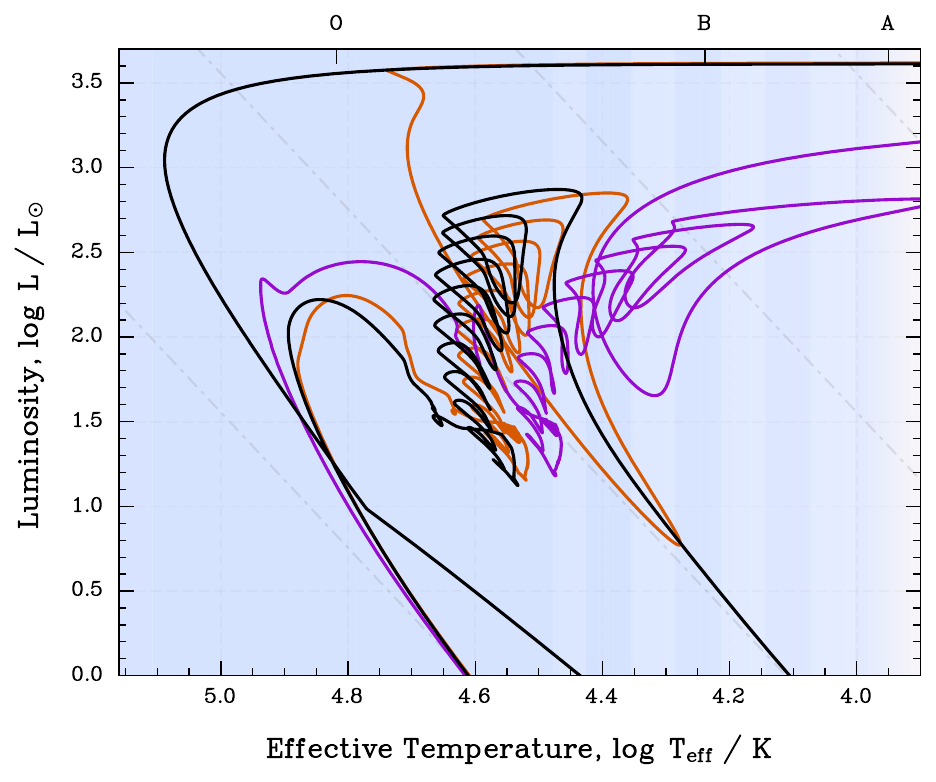}
    \caption{Impact of overshooting prescriptions on the evolution of $0.85~M_\odot$ Pop~III stars. Main-sequence tracks are nearly identical, but enhanced CBM during helium burning enlarges convective cores, yielding higher peak luminosities and modified post-RGB morphologies. {\it First panel}: RGB–DCF phases. {\it Second panel}: EHB and pre-WD stages.}
    \label{fig:OV_CMD}
\end{figure}

\begin{figure*}[t]
    \centering
    \includegraphics[width = \linewidth]{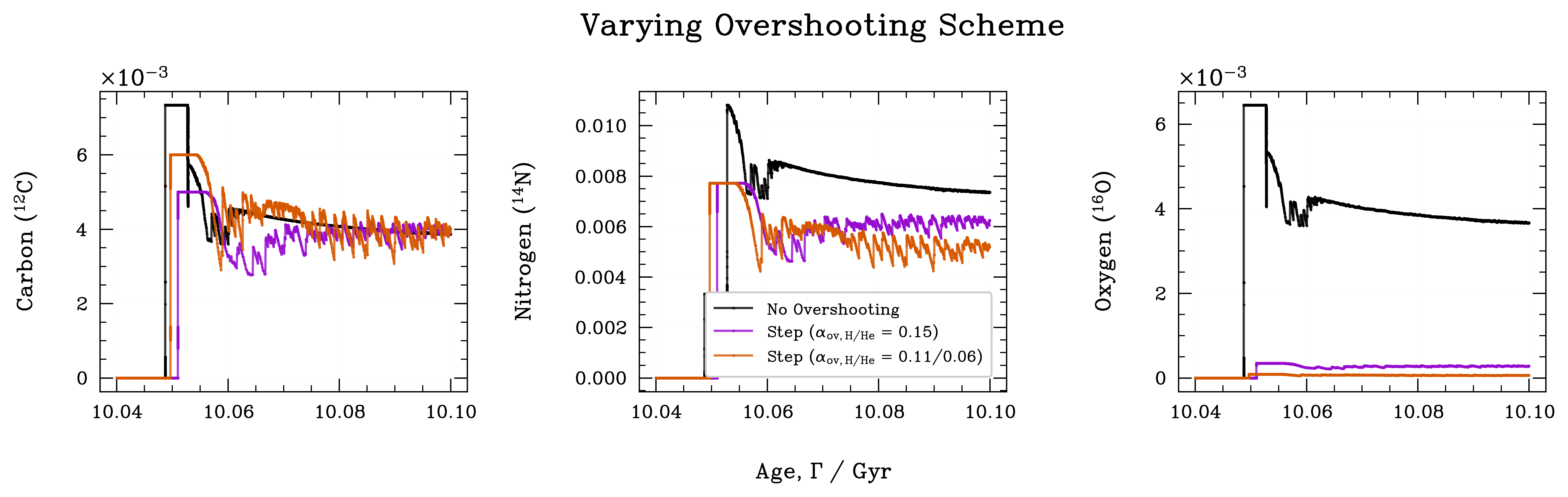}
    \includegraphics[width = \linewidth]{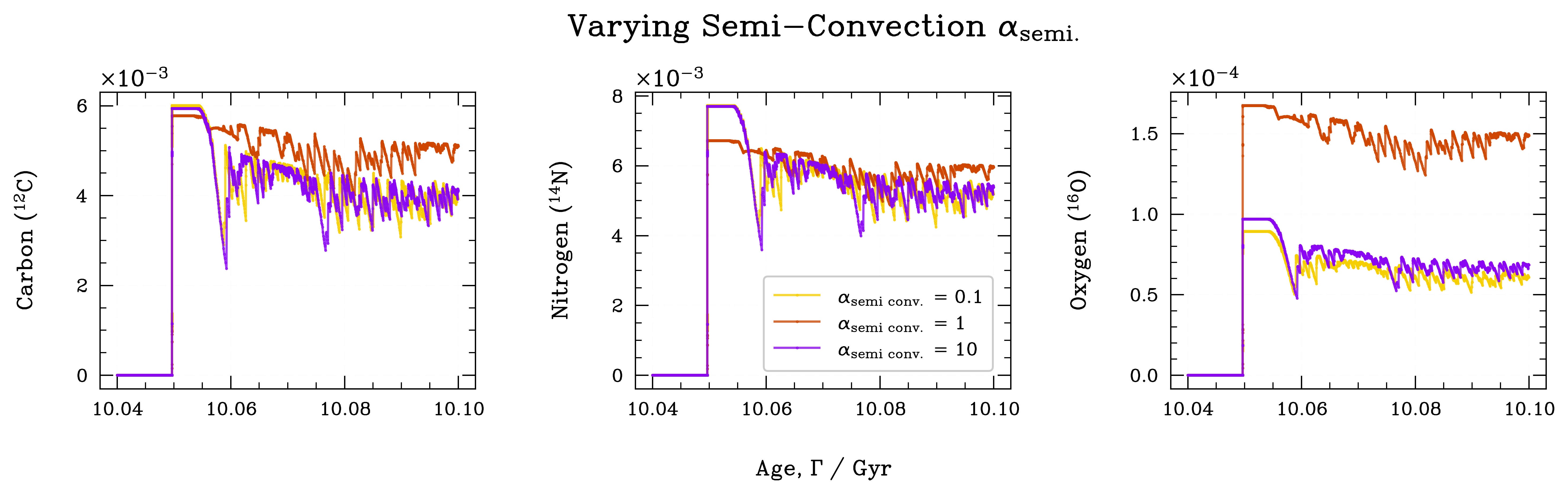}
    \includegraphics[width = \linewidth]{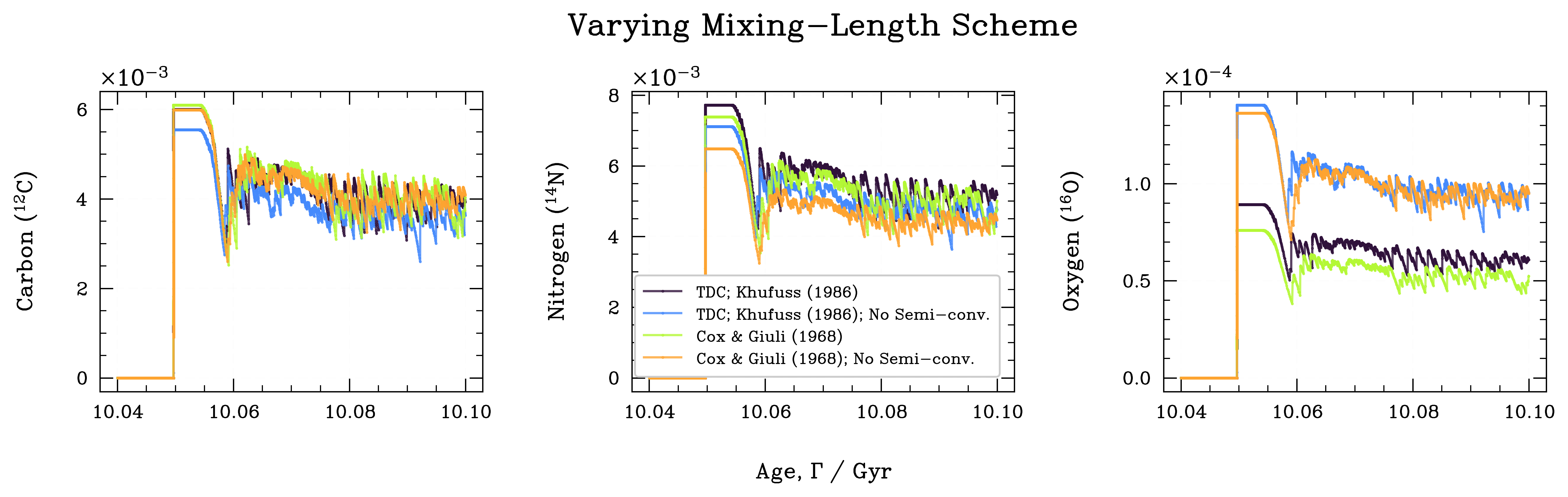}
    \caption{{{\it Top and middle rows}:} Surface CNO abundance evolution during the late {RGB/DCF and early EHB} phases of $0.85~M_\odot$ Pop~III stars. Top row: varying CBM efficiency, showing increasing N/O ratios (1.8 to 87) and modest declines in total CNO abundance with enhanced overshooting. Middle row: varying semi-convection $\alpha_{\rm semi}$. $^{16}$O exhibits pronounced sensitivity, whereas $^{12}$C and $^{14}$N remain largely unchanged. {{\it Bottom row}: Comparison between surface CNO abundances during the late giant branch and HB/EHB phases ($10.04-10.10$ Gyr) for a $0.85~M_\odot$ Pop III star computed with TDC based on \cite{1986A&A...160..116K} model and classical \cite{1968pss..book.....C} MLT, with and without semi-convection. All models converge to similar equilibrium abundances on longer timescales.}}
    \label{fig:OV_nucleo}
\end{figure*}

Figure~\ref{fig:OV_CMD} illustrates the H-R diagram evolution for a $0.85~M_\odot$ Pop~III star under three CBM prescriptions (see details in \S\ref{sec:inputphysics}): (I) no overshooting; (II) uniform step overshooting with the same efficiency for hydrogen- and helium-burning regions; and (III) step overshooting with differentiated efficiencies for hydrogen-burning and helium-burning shells. During the main sequence, all tracks are nearly identical, reflecting the minor role of CBM in low-mass, metal-free stars. Post-MS, differences emerge: model (III) reaches peak luminosities higher by $\Delta\log L \approx 0.1$ dex and follows a distinct sequence of thermal pulses relative to the models (I) and (II), due to enlarged helium-burning convective cores ($\Delta M_{\rm conv}/M_\star \approx 0.02-0.05$), which in turn affects subsequent DCF evolution and, modestly, the final WD mass.

{Finally, we stress that the mixing processes explored in this work are not invoked as a final explanation for the origin of surface CNO enhancements. In low-mass Pop III stars, significant changes in envelope composition naturally occur during advanced evolutionary phases as the convective envelope deepens and engulfs layers previously processed by nuclear burning; observationally, such signatures are predominantly detected on the upper RGB and later stages (e.g., \citealt{2005A&A...430..655S}). Consequently, these abundance patterns may be interpreted as the outcome of standard dredge-up acting on an evolving internal structure, rather than as evidence for additional mixing during earlier phases or for chemical inheritance from the primordial progenitor. Our exploration of semi-convection and CBM serves primarily to quantify structural and evolutionary sensitivities, and to identify the conditions under which non-standard mixing prescriptions materially influence the stellar outcome.} {We tested our combined use of TDC with overshooting and semi-convection by comparing giant branch and early HB/EHB CNO evolution against \cite{1968pss..book.....C} MLT prescriptions, finding that, whilst TDC captures short-timescale mixing variability, all models converge to similar equilibrium abundances on longer timescales as seen in Figure~\ref{fig:OV_nucleo}.}

The nucleosynthetic consequences of CBM are evident in Figure~\ref{fig:OV_nucleo}, which shows surface CNO abundances during the late RGB phase ($10.04-10.10$ Gyr). First dredge-up brings core-processed material to the surface, with the depth of mixing strongly dependent on overshooting efficiency. Without overshooting (model I), surface nitrogen is moderately enhanced (${\rm N/O} \approx 1.8$) with balanced CNO ratios. Uniform step overshooting (model II) produces deeper mixing, yielding extreme oxygen depletion (${\rm N/O} \approx 22$), whilst differential-efficiency overshooting (model III) drives near-total oxygen consumption (${\rm N/O} \approx 87$). The ON sub-cycle of the CNO bi-cycle, $^{16}{\rm O}({\rm p},\gamma)^{17}{\rm F}(\beta^+\nu)^{17}{\rm O}({\rm p},\alpha)^{14}{\rm N}$, is highly efficient at $T \gtrsim 3\times 10^7~K$, and enhanced CBM prolongs residence times in the convective core, facilitating extreme depletion. We notice that the total surface CNO abundance decreases modestly with increasing overshooting ($\Delta{\rm [CNO]} \approx 0.006$ dex).

These nucleosynthetic signatures have major implications for early Galactic chemical evolution. Pop~III stars with substantial CBM produce markedly different CNO yields, with extreme ${\rm N/O}$ ratios ($\gtrsim 20$) consistent with some ultra–metal-poor stars (e.g., \citealt{2005A&A...430..655S}). If such stars inherited their abundances from Pop~III progenitors, accurate CBM modelling is essential to reproducing their chemical fingerprints. Enhanced CBM also reduces total CNO yields, implying that surveys targeting Pop~III relics must account for overshooting when constraining the primordial IMF and early star formation. The factor-of-fifty variation in ${\rm N/O}$ across plausible CBM prescriptions highlights the importance of robust calibration through multi-dimensional hydrodynamic simulations, potentially supported by asteroseismic constraints from upcoming space missions.

{We highlight in this case the findings of earlier Pop III calculations by \cite{2023MNRAS.525.4700L}, whilst noting that the immediate post-DCF evolution depends sensitively on the connection between structural readjustment and envelope mixing (see their Figures $1-5$). In models without convective boundary mixing, the initial response is dominated by core expansion and increasing luminosity, with the effective temperature remaining nearly constant until CNO-enriched layers are later incorporated into the convective envelope. When overshooting is included, enhanced mixing briefly favours a more compact envelope, producing a short-lived increase in effective temperature before opacity-driven expansion sets in. Once dredge-up is established, the subsequent redward evolution converges towards the ones described in previous studies.}

\section{Neutrinos from Low-Mass Pop~III Stars}\label{sec:neutrinos}

\begin{figure*}[t]
    \centering
    \includegraphics[width = 0.49\linewidth]{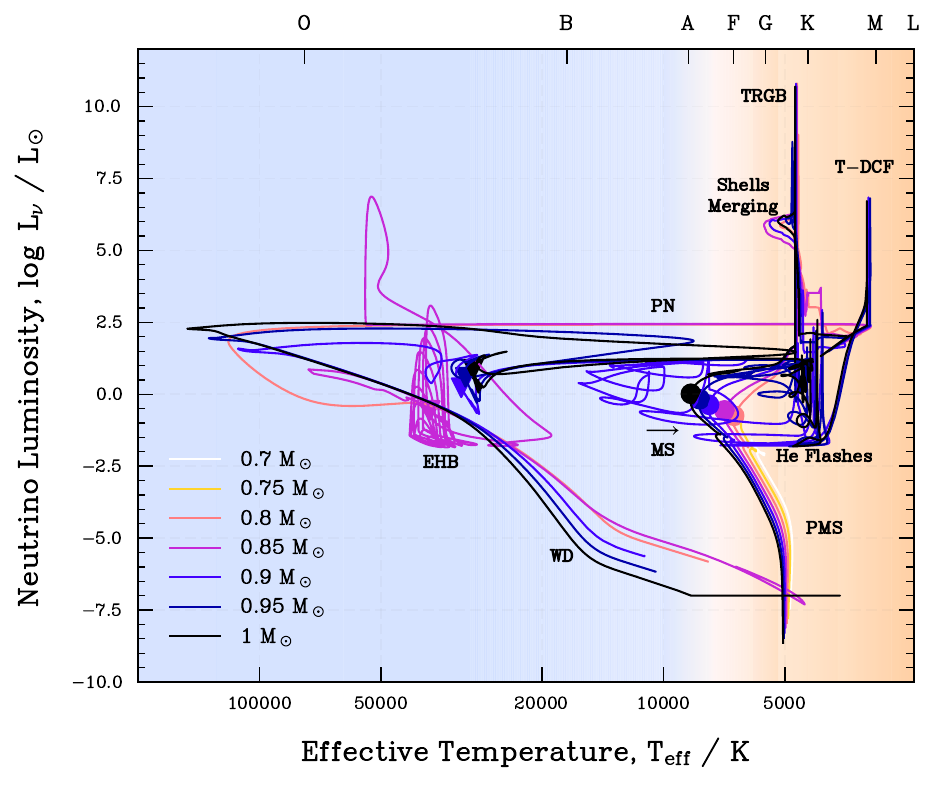}
    \includegraphics[width = 0.49\linewidth]{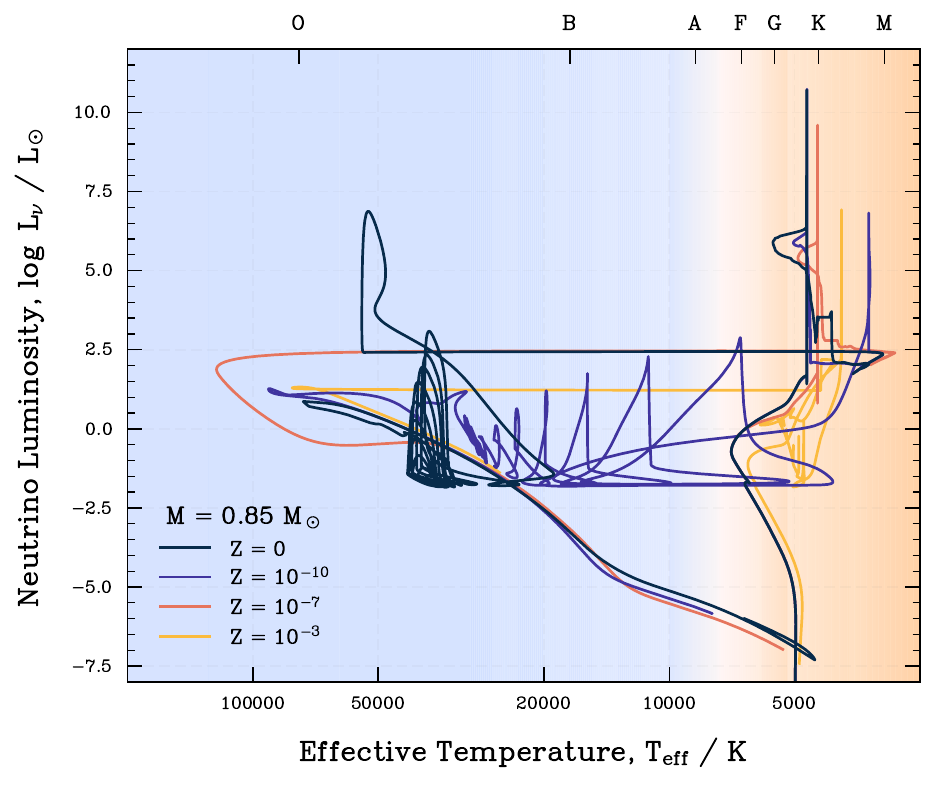}
    \caption{Neutrino H-R diagrams showcasing the evolutionary tracks of stars with masses ranging from 0.7 to 1.0 $M_\odot$ from the PMS to the WD phase (left) and models of initial mass $0.85~M_\odot$ varying metallicity (right). Background colours indicate spectral types based on effective temperature. Filled circles indicate the point where the central hydrogen mass fraction drops below 0.01 ($X_c < 0.01$), whilst inverted triangles mark when the central helium mass fraction drops below 0.01 ($Y_c < 0.01$). During the initial hydrogen-burning phase, neutrino production is dominated by pp-chain reactions and the CNO cycle. This phase corresponds to a plateau in neutrino luminosity due to continuous energy generation through weak interactions. As Pop~III stars transition to helium burning, they undergo the helium core flash, a phase marked by intense bursts of neutrino luminosity primarily driven by the $^{18}{\rm F}\left(,\beta^{+} + \nu_e\right)^{18}{\rm O}$ reaction. Such characteristic spikes in neutrino luminosity are robust indicators of helium ignition and subsequent stable helium burning.}
    \label{fig:neutrino}
\end{figure*}

No low- or high-mass Pop~III stars have been directly detected. Nevertheless, models suggest that low-mass Pop~III stars could persist locally or at high redshift, though their extreme faintness and distance would require strong lensing magnification for detection \citep{2018ApJS..234...41W, 2022MNRAS.512.3030V, 2024MNRAS.533.2727Z}. Additional complications arise from dust attenuation and cosmological dimming, but weakly interacting particles such as neutrinos may escape these obstacles, offering alternative probes of Pop~III stars. The neutrino H-R diagrams ($\nu$H-R) for low-mass Pop~III models, as well as models varying metallicities for $0.85~M_\odot$ models, are shown in Figure~\ref{fig:neutrino}, revealing distinct and dynamic neutrino emission patterns relative to solar-metallicity models (e.g., \citealt{2020ApJ...893..133F, 2024ApJS..270....5F}). 

A striking feature of the Pop~III $\nu$H-R diagram is the presence of sharp, transient peaks in neutrino luminosity, reaching values up to $\log({\rm L}_\nu/{\rm L}_\odot) \approx 7-10$ at lower effective temperatures during the RGB and DCF phases, reflecting brief and intense episodes of nuclear burning plus structural reconfigurations within a star. For instance, in a $0.85~M_\odot$ star model, shell mergers during the TRGB phase can produce neutrino bursts\footnote{During the TRGB phase, shell mergers occur when the hydrogen- and helium-burning shells interact, causing a rapid rise in local temperature and density. This may trigger enhanced weak interaction rates---primarily electron captures and beta decays (e.g., $^{13}\mathrm{N} \rightarrow {}^{13}\mathrm{C} + e^+ + \nu_e$, $^{15}\mathrm{O} \rightarrow {}^{15}\mathrm{N} + e^+ + \nu_e$)---which efficiently produce neutrinos. The electron-degenerate core allows for rapid energy release, resulting in a brief but intense neutrino burst as the star readjusts to a new equilibrium.} with $\log({\rm L}_\nu/{\rm L}_\odot) \approx 10.7$, through these last only about one day; Comparable peaks and durations are seen in higher-mass models. Later, off-centre helium flashes generate further bursts, with $\log({\rm L}_\nu/{\rm L}_\odot) \approx 6.8$ for a $0.85~M_\odot$ Pop~III star, persisting for roughly 600 days\footnote{Off-centre helium flashes occur when the degenerate helium core ignites under conditions of partial electron degeneracy, typically away from the very centre due to the temperature profile. The resulting thermonuclear runaway drives rapid $3\alpha$ reactions and subsequent $\alpha$-captures, producing unstable isotopes that undergo beta decay (e.g., $^{18}\mathrm{F} \rightarrow {}^{18}\mathrm{O} + e^+ + \nu_e$). These weak interaction processes generate sustained neutrino emission. The extended duration (hundreds of days) reflects the timescale over which the core gradually lifts degeneracy and transitions to stable helium burning (see also \S\ref{sec:evolutionary_tracks}).}.

Although the peak neutrino luminosities produced during shell mergers in these low-mass Pop~III stars can reach $\sim1.9-2.5 \times 10^{44}$ erg s$^{-1}$---exceeding the photon luminosity by a factor of $10^{8}$ and the helium-burning luminosity by about $100$---they remain orders of magnitude below the outputs of the brightest extragalactic transients, e.g., {gamma-ray bursts (GRBs), $\approx10^{49}-10^{53}$ erg s$^{-1}$} \citep{1997PhRvL..78.2292W, 2023A&A...672A.102L}; super-luminous supernovae (SLSNe), $\approx10^{54}$ erg s$^{-1}$ \citep{2018SSRv..214...59M, 2024PJAB..100..190Y}; tidal disruption events (TDEs), $\approx 10^{44}-10^{48}$ erg s$^{-1}$ \citep{2024ApJ...970L...8Y, 2024PhRvD.110d3029W}; and active galactic nuclei (AGNs), $10^{43}$–$10^{48}$ erg s$^{-1}$ \citep{2021ApJ...917L..28Z, 2025ApJ...984...54K}. Similarly, the neutrino bursts associated with helium flashes, whilst still dominant (with neutrino luminosity exceeding photon luminosity by $\sim2 \times 10^{3}$ and is comparable to or slightly greater than the helium-burning luminosity), are less extreme, peaking at $\sim2 \times 10^{40}$ erg s$^{-1}$. Therefore, whilst these events represent some of the most energetic neutrino phenomena in stellar evolution, their luminosities are modest compared to the brightest extragalactic transients and distinguishing among them would be complicated.

The metallicity-dependent neutrino emission patterns in $0.85~M_\odot$ models reveal fundamental differences in nuclear reaction networks and energy transport. In Pop~III stars, the initial absence of CNO catalysts forces hydrogen burning to proceed predominantly via the pp chain, producing electron neutrinos through weak interactions such as $p(p,e^+\nu_e)d$. CNO burning only becomes significant during the TRGB phase, when shell mergers facilitate deeper mixing, enhancing $\beta^{+}-$decay processes in the CNO cycle---particularly $^{13}{\rm N}(\beta^+\nu_e){}^{13}{\rm C}$ and $^{15}{\rm O}(\beta^+\nu_e){}^{15}{\rm N}$---and producing neutrinos with higher energies. In contrast, metal-rich stars sustain efficient CNO cycling from the main sequence \citep{1967ApJ...149..117D}. During helium flashes, the neutrino luminosity spikes ($\log(L_\nu/L_\odot) \approx 6.8$) as a result of electron captures and $\beta^{+}-$decays of unstable isotopes synthesised through $\alpha-$capture reactions, notably $^{18}{\rm F}(\beta^+\nu_e){}^{18}{\rm O}$, under conditions of partial electron degeneracy ($\rho \sim 10^6~{\rm g~cm^{-3}}$, $T \sim 10^8~{\rm K}$). The elevated electron chemical potential $\mu_e$ enhances weak interaction rates via Fermi–Dirac statistics, producing neutrino energy spectra and emission timescales that are distinct from those of metal-rich stars.

\begin{figure}[t]
    \centering
    \includegraphics[width = \linewidth]{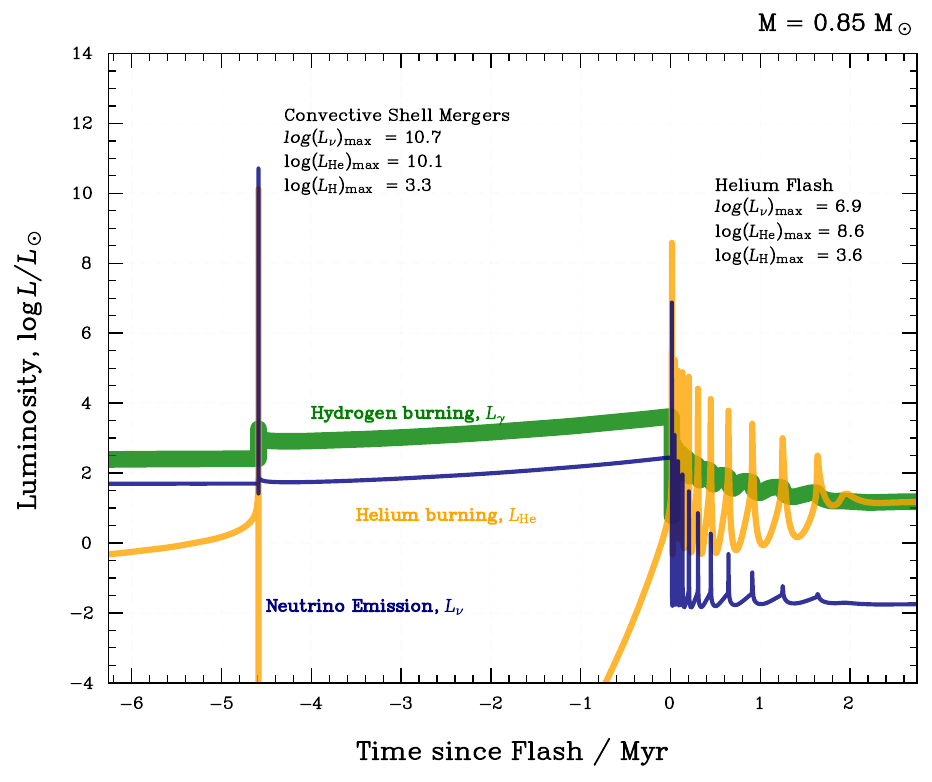}
    \caption{Temporal evolution of hydrogen burning luminosity (${\rm L}_{\gamma, \rm H}$), helium burning luminosity (${\rm L}_{\rm He}$), and total neutrino luminosity (thermal plus nuclear; ${\rm L}_{\nu}$) for a $0.85~M_\odot$ Pop~III model. Similar behaviours are seen for the higher masses ($0.9-1.0~M_\odot$), with slight differences in the luminosity peak and duration of the flashes. The abscissa quantifies the time elapsed since the off-centre helium flash in million years. Shell mergers during the RGB phase and subsequent off-centre helium flashes' peak luminosities are indicated.}
    \label{fig:neutrino_luminosities085}
\end{figure}

Figure~\ref{fig:neutrino_luminosities085} depicts the evolution during helium burning of key energy generation and loss channels for a $0.85~M_\odot$ Pop~III model. The time axis is referenced to the onset of the off-centre helium flash, as discussed in Sections \ref{sec:08}-\ref{sec:095-1}. The hydrogen and helium burning luminosities show the transition from core hydrogen burning to shell burning as the star evolves. Notably, the neutrino luminosity exhibits two sharp peaks: the first, at $\Gamma \approx 10.05$, coincides with {the DCF phase} leading to a shell merger event, which temporarily disrupts the model and triggers a strong CNO flash. The second, approximately after $\approx$4~Myr, marks a standard core helium flash.

Beyond the total luminosity of these explosive events, the detailed characteristics (flavours, masses, and time-travels) of neutrino emission throughout the evolution of low-mass Pop~III stars are critical. Their pristine composition leads to distinct internal conditions and nuclear-burning pathways compared to metal-rich stars, which may result in unique neutrino energy spectra and timing profiles. For example, the relative contributions from the pp chain versus the CNO cycle, or specific reactions during helium flashes and thermal pulses, can shape both the energy and temporal distribution of emitted neutrinos during the MS, RGB and DCF phases. In these cases, specific core and shell conditions---such as temperature, density, and electron fraction---directly affect the production rates of different neutrino flavours (either electron $\nu_e$, muon $\nu_\mu$, and tau $\nu_\tau$). Neutrino production in low-mass Pop~III stars may primarily occur via nuclear fusion and weak interaction processes in their interiors. Unlike massive stars or CCSNe, the temperatures and densities in these stars are generally too low to produce significant numbers of muon or tau neutrinos. Such flavours are typically generated in more extreme environments, like collapsing massive stars or through hadron decays---specifically from Pi meson and K meson---in AGN or GRBs \citep{2022arXiv220206480K, 2025ApJ...982...94P}. For context, decays such as $\pi^{+} \rightarrow \mu^{+} + \nu_\mu$, ${\rm K}^{+} \rightarrow \mu^{+} + \nu_\mu$, followed by $\mu^{+} \rightarrow e^{+} + \nu_e + \overline{\nu_\mu}$, are important sources of muon and tau neutrinos in these environments. In Pop~III stars, however, the dominant neutrino flavours may be electron neutrinos ($\nu_e$) and electron antineutrinos ($\overline{\nu}_e$), produced mainly through the pp chain and subsequent beta decays (e.g., $p+p \rightarrow {}^{2}\mathrm{H} + e^{+} + \nu_e$, $^{8}\mathrm{B} \rightarrow {}^{8}\mathrm{Be}^* + e^{+} + \nu_e$, $n \rightarrow p + e^{-} + \overline{\nu}_e$), and, if heavy elements are dredged up, from the CNO cycle ($^{13}\mathrm{N} \rightarrow {}^{13}\mathrm{C} + e^{+} + \nu_e$, $^{15}\mathrm{O} \rightarrow {}^{15}\mathrm{N} + e^{+} + \nu_e$). The metal-free composition of Pop~III stars means the CNO cycle is absent or highly suppressed until shell mergers at the TRGB, forcing hydrogen burning to proceed via the pp chain. The resulting flavour ratios, modified by oscillations/superpositions over cosmological distances, could provide unique signatures for identifying Pop~III neutrino sources, although the detailed physics behind these phenomena is outside the scope of this work and could be explored in further dedicated studies (e.g., \citealt{2008ApJ...675..937I}). 

{Current low-energy neutrino observatories, such as Borexino \citep{2022PhRvL.128i1803A, 2023PhRvD.108j2005B, 2024ARNPS..74..369B}, have demonstrated the ability to detect solar neutrinos down to sub-megaelectronvolt energies via elastic $\nu_e-e^{-}$ scattering in large volume liquid scintillators, with recent reports of $^{8}{\rm B}~\longrightarrow~^{8}{\rm Be} + e^{+} + \nu_e$ neutrinos above 3 MeV \citep{2017arXiv170900756T}, highlighting the potential for observing low-energy fluxes when background noise are sufficiently suppressed. Despite such remarkable sensitivity, the transient neutrino bursts predicted for low-mass Pop III stars in this work may pose a distinct challenge, as shell produce short-duration ($\sim$minutes to hours; see Figure \ref{fig:neutrino_luminosities085}) spikes in neutrino luminosity, which would generate only a handful of interactions even if the source were nearby, whilst cosmological distances reduce the expected flux by many orders of magnitude. Consequently, a direct detection with current instruments is effectively precluded. Isolating such signals would require next-generation detectors combining very large ``target masses" (i.e., mass of material within the detector that neutrinos can interact with), low intrinsic background, and high temporal resolution. Scintillation-based observatories with sub-second timing could, in principle, resolve the brief bursts, whilst multi-detector coincidence could distinguish rare astrophysical events from local noise. Energy reconstruction would further aid discrimination, as Pop III bursts may be dominated by $e^{-}-$neutrinos from pp-chain and CNO reactions, with characteristic spectra distinct from solar and/or reactor backgrounds. Whilst these capabilities remain conceptual, they provide a framework for identifying neutrino signatures of low-mass, metal-free stars, extending the diagnostic power of neutrino astronomy beyond local and continuous sources.}

\section{Discussions and Conclusions}\label{sec:conclusions}

We presented evolutionary models for low-mass Pop~III stars ($0.7-1.0~M_\odot$) evolved to the WD phase using the MESA stellar evolution code. Details are presented in Appendix \ref{app:evolution}. Our results highlight fundamental differences between metal-free stars and their extremely metal-poor or solar-metallicity counterparts beyond spectroscopic properties, with implications for evolutionary pathways, seismic diagnostics, and potential detectability both locally and at high redshift. Several aspects of our models remain, however, subject to physical uncertainties. In particular, the occurrence of shell mergers during the red-giant branch (RGB) is highly sensitive to the adopted treatment of convective overshooting. This phenomenon, also reported in previous studies (e.g., \citealt{1990ApJ...349..580F, 1996ApJ...459..298C, 2002A&A...395...77S, 2004A&A...422..217W, 2023MNRAS.525.4700L}), may not be robust across all overshooting prescriptions, and its interpretation requires caution. Likewise, mass-loss rates on the giant branches for metal-free stars are still poorly constrained (cf. Figure~7 in \citealt{2025arXiv250512794L} and \citealt{2001ApJ...559.1082S, 2008A&A...490..769C}), and deviations from standard prescriptions can introduce uncertainties in late-stage evolution and final remnant masses.

\begin{enumerate}
    
    \item {\it Thermal Compactness and Enhanced Radiative Stratification}. Due to their primordial composition, Pop~III stars exhibit significantly reduced radiative opacities, particularly in the outer and partially ionised zones where metals usually dominate bound-bound and bound-free opacity sources. This results in highly efficient radiative energy transport and relatively shallow radiative temperature gradients, leading to compact, centrally condensed stellar structures with high central temperatures. The absence of metal-driven opacity buffering also inhibits the development of deep convective envelopes, reinforcing strongly stratified, radiative interiors through much of the star's lifetime (e.g., Figures \ref{fig:kipp085}, \ref{fig:kipp08}, \ref{fig:kipp09}, and \ref{fig:kipp095-1}). \smallskip

    \item {\it Extended Main Sequence Lifetimes and Cosmological Survivability}. The main sequence lifetimes of low-mass Pop~III stars follow a steep mass-luminosity-age scaling relation of the form $\Gamma_{\rm MS} \propto {\rm M}^{-3.2}$, significantly steeper than the canonical $M^{-2.5}$ relation for solar-metallicity stars (cf., Figure~\ref{fig:MSlifetimes}). Consequently, Pop~III stars with masses below $0.8~M_\odot$ may still be present in the Milky Way and/or its satellite dwarfs today (e.g., \citealt{1996ApJ...473L..95U, 1999ApJ...515..239N, 2014ApJ...785...73S, 2014ApJ...792...32S, 2015MNRAS.448..568H, 2020ApJ...901...16D}, and references therein). Their detection, however, may be hindered by their low metallicity, small radii, low luminosities, and the potential contamination of their atmospheres via accreted metals. \smallskip

    \item {\it Surface Composition Evolution and Chemical Feedback}. Although Pop~III stars lack initial metals, late-stage dredge-up events driven by shell mergers and thermal instabilities (particularly at the T-RGB and post-DCF stages) {provide significant observable abundance signatures resembling Pop II stars at carbon and nitrogen levels of ${\rm X}_{\rm CNO} \sim 10^{-2}$ (see Figure~\ref{fig:085ejected}).} These final abundances are, however, sensitive to mixing and overshooting schemes (see \S\ref{sec:MLT}, and \S\ref{sec:semiconv}), although no unique solution exists. Without overshooting, the envelope only reaches the outer intershell layers, which are relatively richer in $^{16}{\rm O}$, whereas overshooting digs into deeper, $^{12}{\rm C}-$rich layers, reducing the $^{16}{\rm O}$ fraction (see Figure \ref{fig:OV_nucleo}). The integrated yields of CNO isotopes, whilst modest ($\lesssim 10^{-3}~{\rm M}_\odot$), are not zero, and may contribute to early chemical enrichment of low-mass halos, and may imprint observable abundance signatures in second-generation stars. However, the absence of iron-group and $\alpha-$capture products reflects the lack of core-collapse supernovae (CCSNe) ejecta in this mass regime. \smallskip

    \item {\it Metallicity-Dependency}. The introduction of even trace amounts of heavy elements into low-mass stellar models induces a suite of interdependent structural and evolutionary effects, fundamentally altering energy generation, convective mixing, and envelope behaviour (see \S\ref{sec:comparison_highZ}). Metallicity modifies opacity, which in turn changes the radiative gradient and convective morphology, reshaping the nuclear burning environment. As revealed in Figures~\ref{fig:multi_metal} and \ref{fig:kippmetals}, metal-free ($Z=0$) stars are distinguished by a lack of envelope convection during the RGB phase, a predominantly radiative main sequence interior, violent off-centre helium flashes, and strong DCF pulses. In contrast, even minimal enrichment ($Z\gtrsim10^{-10}$) enables the CNO cycle, triggers convective cores, and produces milder shell instabilities and envelope responses. This feedback shapes the compactness, composition, and cooling behaviour of the final WD remnant, with Pop~III models yielding hotter, more compact WDs with thinner hydrogen envelopes and faster cooling rates. The absence of early dredge-up delays surface pollution until the final DCF pulses, meaning the chemical signature of internal nucleosynthesis is imprinted only very late. If such WDs are ever observed in ancient, metal-poor environments, they would serve as key probes of primordial stellar evolution, late-stage mixing physics, and mass-loss processes in the metal-free regime. \smallskip

    \item {\it Horizontal Branch Evolution and Formation of Helium- and Carbon/Oxygen-core White Dwarfs}. Although metal-free stars are theoretically expected to experience weak radiative winds, our models incorporate enhanced mass loss via Reimers and Bl{\"o}cker prescriptions once surface CNO enrichment occurs, following past studies. We explored different outcomes varying the efficiency of mass loss during the RGB in \S\ref{sec:mass-loss}. During the T-DCF phase, dredging carbon and nitrogen into the envelope increases local opacity, triggering pulsation-enhanced mass loss and enabling substantial envelope stripping. Consequently, stars in the upper mass range ($0.85-1~M_\odot$) experience degenerate helium ignition under off-centre conditions, leading to brief excursions through the EHB stage following the thermal pulsating DCF (see Figures \ref{fig:HR}, \ref{fig:TcRhoC}, and \ref{fig:neutrino_luminosities085}). These phases are marked by short-lived UV-bright episodes with $T_{\rm \star, eff} \sim 50~000-60~000~K$ and $L \gtrsim 300~L_\odot$, which may influence the far-UV component of ancient stellar populations and serve as indirect tracers of Pop~III activity. Their remnants evolve into low-mass (He or CO) WDs with typical final masses of ${\rm M}_{\rm WD} \sim 0.45-0.55~M_\odot$. By Hubble time, low-mass Pop~III WDs may exist today in the Milky Way halo and globular clusters, and these objects, whilst spectroscopically indistinct due to surface pollution and cooling times, may retain internal structural signatures of their metal-free origins---particularly in their mass-radius relations, cooling sequences, and potential seismic behaviour. The properties, detectability, and observational discriminants of these primordial WD remnants will be explored in a forthcoming study. \smallskip

    \item {\it Neutrino Emission from Low-Mass Pop~III Stars}. Low-mass Pop~III stars may produce characteristic neutrino signatures shaped by their pristine, metal-free composition and unique nuclear burning pathways (see \S\ref{sec:neutrinos}). Intense but brief neutrino bursts occur during shell mergers and off-centre helium flashes, reaching peak neutrino luminosities up to $\sim 10^{44}$ erg s$^{-1}$---exceeding photon luminosity by factors of $\sim10^{8}$---yet remaining orders of magnitude below the some of the brightest extragalactic transients ($\sim 10^{53}$ erg s$^{-1}$; see also Figure~\ref{fig:neutrino_luminosities085}). The dominant neutrino flavours could be electron neutrinos and anti-neutrinos from pp-chain and beta decay processes, with minimal contributions from muon and tau neutrinos due to relatively low core temperatures and densities. The suppression of the CNO cycle until late evolutionary phases further shapes neutrino production, which may result in distinct energy spectra and temporal profiles. These features, together with neutrino flavour oscillations over cosmological distances, offer an additional---though challenging and beyond the scope of this work---observational window into primordial populations that are otherwise inaccessible by electromagnetic means, motivating future dedicated studies on neutrino-based probes of the early Universe. {The neutrino bursts associated with shell mergers and helium flashes in low-mass Pop~III stars, however, whilst energetic relative to the star's photon output, should be recognised as extremely brief, occurring on timescales of minutes to days. Given these short durations and the expected distances of Pop~III stars (e.g., \citealt{2018MNRAS.473.5308M}), the direct detection of such events with current neutrino observatories may be highly unlikely with current instruments.} \smallskip

    \item {\it The Pop~III/II Threshold}. The concept of a critical metallicity, $Z_{\rm crit}$, marking the transition from Pop~III to Pop~II star formation, was first introduced in studies investigating the role of metal-line and dust cooling in gas fragmentation \citep{2003ApJ...596...34B, 2003Natur.422..869S}. Subsequent studies also refer to extremely metal-poor stars ($Z/Z_\odot \lesssim 10^{-7}-10^{-4}$) as Pop~III stars (e.g., \citealt{2016ApJ...831..204R, 2017MNRAS.466.4826K, 2018MNRAS.475.4396J, 2019MNRAS.485.5939J, 2022A&A...668A.191C, 2020MNRAS.491.4387K, 2023MNRAS.524..351K}). However, it has become increasingly clear that no unique or universal threshold governs this transition. Instead, it appears to depend sensitively on local conditions, such as gas density, turbulence, and the specific cooling mechanisms at play. The most iron-poor stars discovered to date are SDSS J0715-7334 ($Z \leq 7.8\times10^{-7}$; \citealt{2025arXiv250921643J}), and SDSS J102915.14+172927.9 ($Z \leq 1.9\times10^{-6}$; \citealt{2024A&A...691A.245C}), and despite their extremely low metallicity, both are considered second-generation objects, having formed from gas enriched by the first stars, rather than a true Pop~III star. To date, no genuinely metal-free object has been identified, either of high- or low-mass. 
    
\end{enumerate}

Even modest self-enrichment during stellar evolution can blur the distinction between Pop~III and extremely metal-poor Pop~II stars, complicating efforts to identify truly primordial objects. Surface abundances may be altered not only by accretion from the ISM but also by internal convective dredge-up, particularly following hydrogen-shell merger flashes, such that the photosphere may not faithfully reflect the star’s initial metallicity. As a result, conventional spectroscopic diagnostics alone cannot unambiguously distinguish first- from second-generation stars (see, e.g., \citealt{1986A&A...168...81C}). Probing below the surface is therefore essential, and asteroseismology offers the most promising avenue to reveal the internal structure and composition of candidate Pop~III survivors, as explored in the second paper of this series (Ferreira et al., {\it submitted}). 

\begin{acknowledgments} 

    We warmly thank Sarbani Basu, Pratik Gandhi, and Saskia Hekker for their insightful discussions, which have greatly enriched this work. 
    {We also warmly thank the referee for their thoughtful and constructive comments, which have greatly improved the clarity and depth of this work.}
    This work is dedicated to the memory of Pierre R. Demarque (1932–2025), a pioneer in stellar evolution theory, whose influential papers advanced our understanding of the evolution of the Universe’s earliest low-mass stars. We thank the Yale Center for Research Computing (YCRC) for providing access to the Grace computing cluster used in this study. 
    TF acknowledges support from the Yale Graduate School of Arts and Sciences. 
    EF thanks the Yale Center for Astronomy and Astrophysics Prize Fellowship. 
    CJL acknowledges support from a Gruber Science Fellowship and from NSF grant AST-2205026.

\end{acknowledgments}


\software{MESA \citep{2011ApJS..192....3P, 2013ApJS..208....4P, 2015ApJS..220...15P, 2018ApJS..234...34P, 2019ApJS..243...10P, 2023ApJS..265...15J}.}

\bibliography{bib}{}
\bibliographystyle{aasjournalv7}

\appendix 

\section{The Detailed Evolution of Low-Mass Pop~III stars}\label{app:evolution}

\begin{figure*}[t]
    \centering
    \includegraphics[width = 0.8\linewidth]{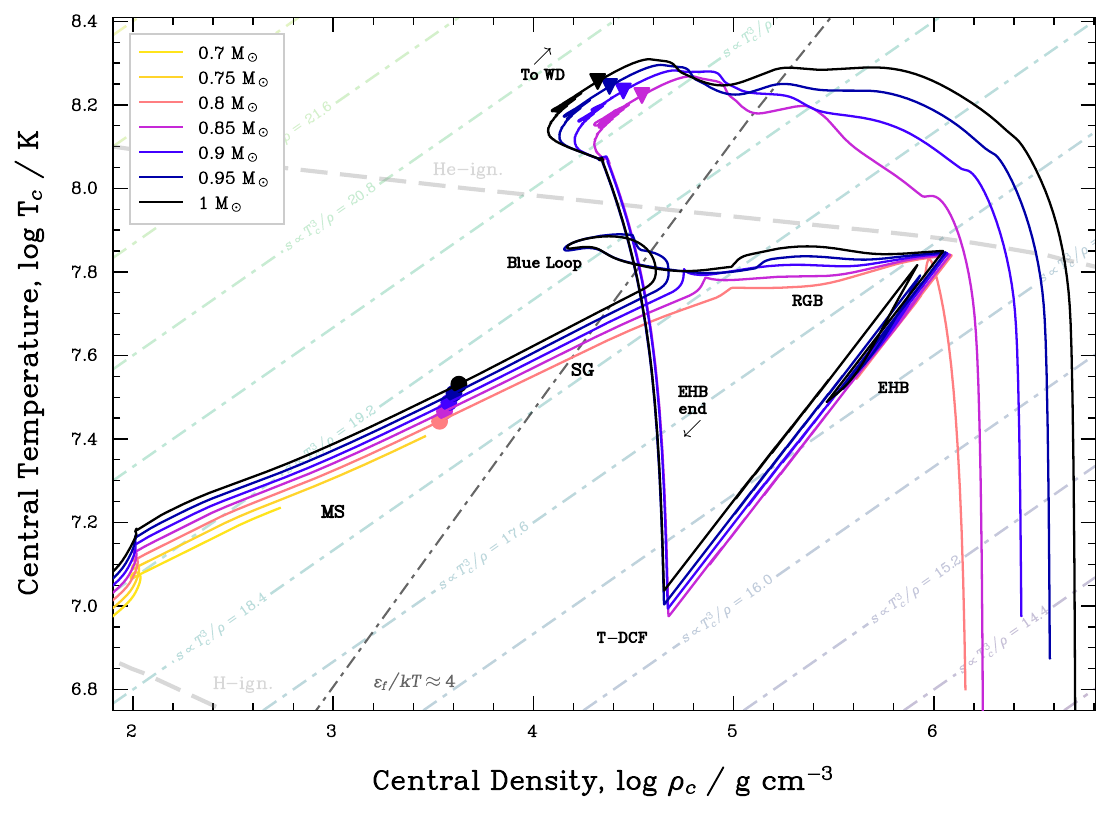}
    \includegraphics[width = 0.8\linewidth]{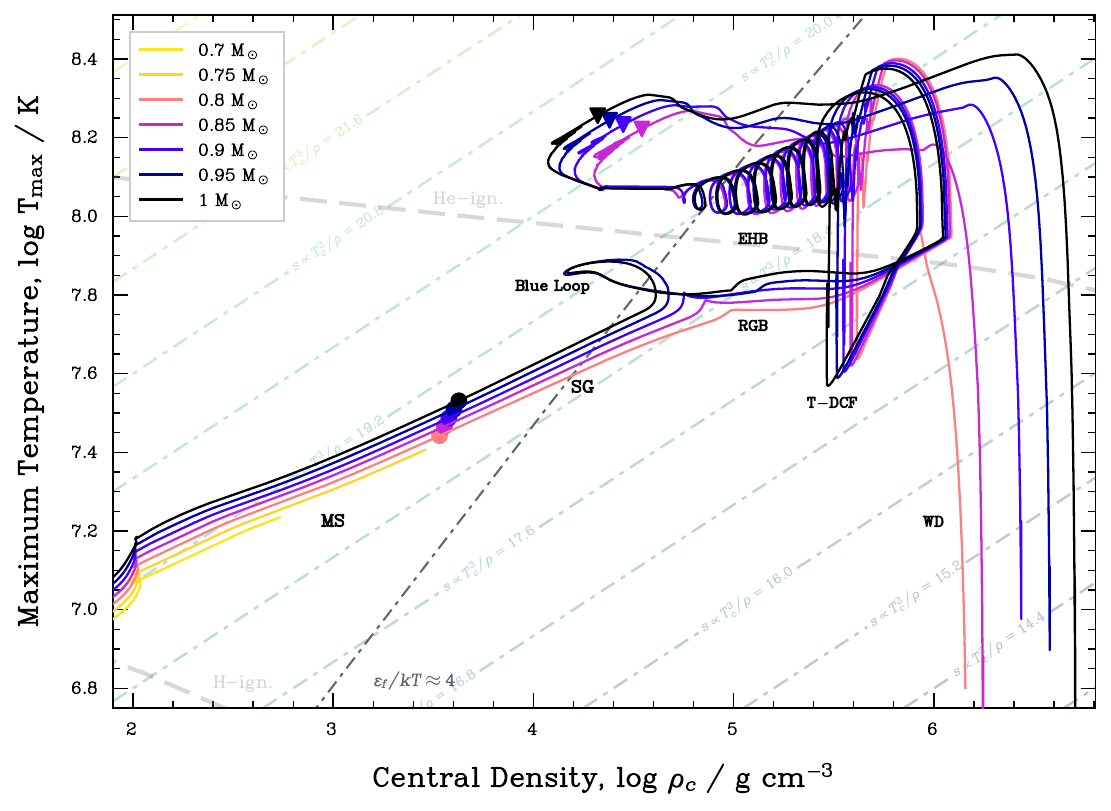}
    \caption{Central temperature ($\log T_{\rm c}$; first panel) and maximum internal temperature ($\log T_{\rm max}$; second panel), along with central density ($\log \rho_{\rm c}$), for Pop~III models with ZAMS masses ranging from $0.7$ to $1.0~M_\odot$. Key nuclear ignition thresholds, including those for hydrogen (H) and helium (He), are shown as dashed lines. The condition under which electron degeneracy pressure becomes significant is indicated based on Fermi energy ($\varepsilon_{\rm F}/\kappa T \approx 4$), whilst the dashed blue line represents the ideal gas regime. Dash-dotted lines denote contours of constant specific entropy proportionality. Filled circles mark the point at which the central hydrogen mass fraction drops below 0.01 ($X_c < 0.01$), and inverted triangles indicate the point where the central helium mass fraction falls below 0.01 ($Y_c < 0.01$). The main evolutionary phases and episodes are noted.}
    \label{fig:TcRhoC}
\end{figure*}

\subsection{$0.7-0.75~M_\odot$ Stellar Models}

These stars do not reach the end of the main sequence within the Hubble time. If they still exist, $0.7-0.75~M_\odot$ Pop~III stars could be identified as relatively hot dwarf main sequence stars ($T_{\rm eff} = 6050-6400~K$) with radii of $R = 0.85-1.3~R_\odot$ by $\Gamma = 13.8$~Gyr of pristine composition. Although their evolution may not be of primary interest, their asteroseismic signatures, when compared to metal-rich counterparts, could be investigated in further work. Energy generation is primarily through the proton–proton (pp) chain. {The stellar interiors are radiative over essentially the entire mass, with no extended convective envelope developing during the main-sequence evolution. Surface convection associated with hydrogen partial ionisation is confined to an extremely shallow outer layer and does not contribute significantly to the global structure.} The surface hydrogen and helium mass fractions evolve to ${\rm X}_{\rm H} = 0.895-0.876$ and ${\rm X}_{\rm He} = 0.104-0.123$, respectively, by $\Gamma = 13.8$ Gyr. {This increase in surface hydrogen abundance and corresponding depletion of helium is driven by gravitational settling operating throughout the main-sequence evolution.}

\subsection{$0.8~M_\odot$ Pop~III Stellar Model}\label{sec:08}

\begin{figure*}[t]
    \centering
    \includegraphics[width = \linewidth]{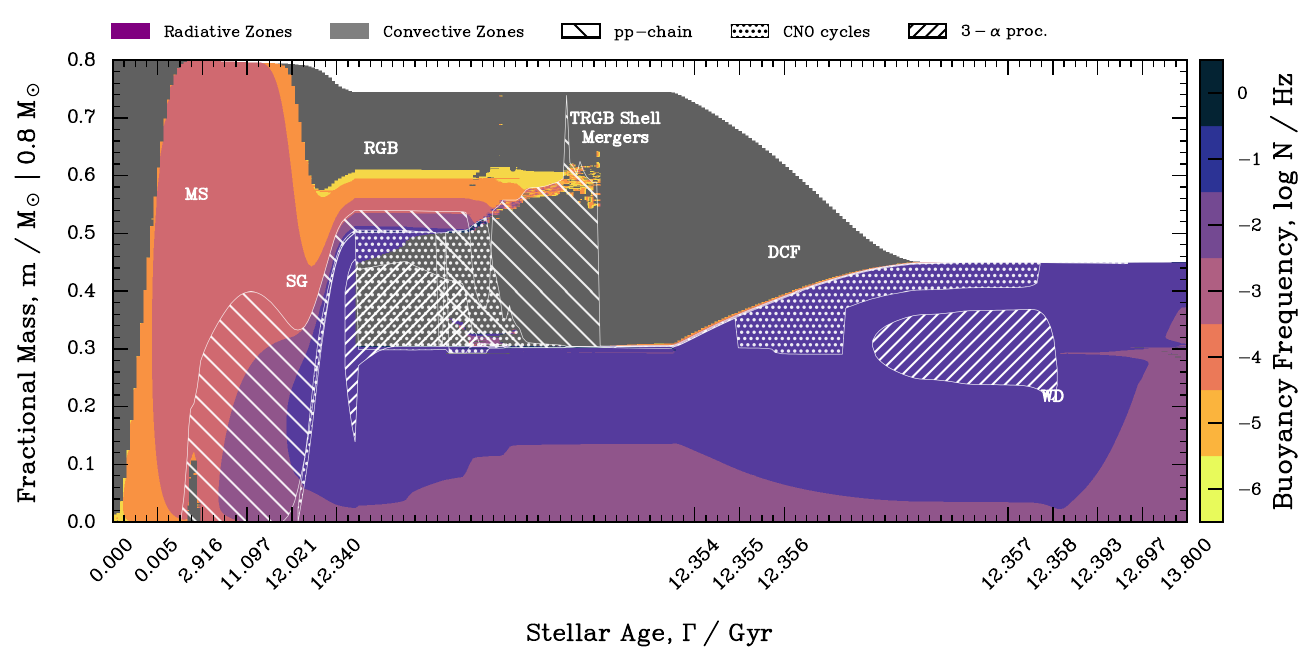}
    \caption{Kippenhahn diagram for a $0.8~M_\odot$ Pop~III star, shown from the PMS to the WD phase. Radiative zones are depicted using a colour scale, whilst convective regions (where the Brunt-V{\"a}is{\"a}l{\"a} frequency satisfies $N^2 < 0$) are shown in grey. Nuclear burning regions are marked using distinct hatching: leftward diagonal lines for the pp-chain, dots for the CNO cycle, and rightward diagonal lines for the $3\alpha$ process.}
    \label{fig:kipp08}
\end{figure*}

During the main sequence (MS), a $0.8~M_\odot$ Pop~III star primarily generates energy via the pp-chain (see Figure~\ref{fig:kipp08}). The core remains radiative, whilst the outer layers develop a shallow convective envelope due to partial ionisation zones and evolving temperature gradients. This convective envelope persists as the star transitions from the MS to the sub-giant (SG) branch. At this phase, the hydrogen-burning shell moves outward, causing the core to contract and heat up, marking the beginning of the star's ascent of the RGB. As the core contracts and temperature increases, the star fuses predominantly helium. During the RGB, energy generation is still primarily due to the pp-chain, with carbon-nitrogen-oxygen (CNO) cycles becoming more important in later phases. 

As the star ascends the RGB, the convective envelope deepens significantly due to the steep temperature gradient in the outer layers. The core, composed mainly of helium ashes, remains radiative, whilst the surrounding envelope undergoes extensive convection. The star's luminosity increases dramatically, as does its radius. 
At the tip of the RGB (T-RGB), the core contracts and heats up due to increased electron degeneracy, whilst a convective region develops above the hydrogen-burning shell and beneath the outer convective envelope. The increasing central degeneracy ($\eta_c > 10$) confirms the highly degenerate state of the helium core, supporting itself via electron degeneracy pressure and preventing the occurrence of a classical helium flash for a $0.8~M_\odot$ Pop~III star. 

The helium core remains inert, whilst burning within the hydrogen shell continues. Energy generation at the T-RGB is still dominated by hydrogen shell burning, with the hot CNO cycle overtaking the pp-chain due to gradual self-enrichment. $3\alpha$ processes are negligible at this phase, with the helium core still inert and degenerate. The onset of $3\alpha$ reactions initiates carbon burning in an inner shell, which subsequently expands outward whilst the core continues to cool and contract. 
This expansion reaches the radiative hydrogen envelope, {triggering the DCF phase (see Figure~\ref{fig:neutrino_luminosities085} and \citealt{2008A&A...490..769C}}). Consequently, significant production of $^{14}$N occurs, as the CNO cycle is now active in the helium shell. The merger of the burning shells leads to convective dredge-up, which propagates a convective wave to the stellar surface. This results in short-lived excursions towards the blue side of the H-R diagram at effective temperatures of $T_{\rm eff} \sim 40~000~K$ and luminosities of $L \sim 430~L_\odot$. At $\Gamma\approx12.3$~Gyr---during the transition to the DCF phase and at a mass of $M\approx0.73~M_\odot$---the surface mass fraction of CNO elements reaches ${\rm X}_{\rm CNO}\sim0.085$ {(see Table \ref{tab:PopIIICNO})}, with carbon (${\rm X}_{^{12}\rm C} \sim 0.0065$) and nitrogen (${\rm X}_{^{14}\rm N} \sim 0.0087$) being the dominant contributors. It thus becomes a nitrogen-rich, ultra-metal-poor star. {We note, however, that this evolutionary pathway is particularly sensitive to the treatment of convective overshooting in hydrodynamical models that rely on the pressure scale height, since the stellar structure is evolved using a time-dependent one-dimensional hydrodynamic scheme, in which finite accelerations and inertia are explicitly accounted for, rather than through a sequence of purely hydrostatic configurations (cf. Section 4 in \citealt{2015ApJS..220...15P}).} If the distance between two burning shells is less than a single pressure scale height ($H$)---influenced by the mean molecular weight ($\mu$), which decreases as the model's metallicity decreases---they will mix (see e.g., \citealt{2024ApJS..270....5F}). 

\begin{deluxetable*}{c|c|c|c}
    \label{tab:PopIIICNO}
    \tablecaption{{Surface CNO enrichment in our Pop~III models compared with values from LM23. $^\dagger$Initial model's mass. The fractional difference $(X_{\rm CNO} - Z_{\rm DCF})/Z_{\rm DCF}$ is denoted by $\Delta$.}}
    \tablehead{
    \colhead{Mass$^\dagger$ (M$_\odot$)} & \colhead{X$_{\rm CNO}$ (This work)} & \colhead{Z$_{\rm DCF}$ (LM23)} & \colhead{$\Delta$}}
    \startdata
    0.8   & 0.085   & 0.0144 & +4.90 \\
    0.85  & 0.013   & 0.0125 & +0.04 \\
    0.9   & 0.0125  & 0.0112 & +0.12 \\
    0.95  & 0.0116  & 0.0102 & +0.14 \\
    1.0   & 0.0107  & 0.0093 & +0.15 \\
    \enddata
\end{deluxetable*}

Further evolving towards the DCF phase, the star's structure consists of an outer convective envelope and no energy production in the highly degenerate core, but instead production in thin shells around the core. The hydrogen- and helium-burning shells operate quasi-independently, separated by a thin radiative layer, and exhibit mid-thermal instabilities but no large-scale helium shell flashes. By reaching the tip of the DCF (T-DCF), a tiny radiative region in mass beneath the convective envelope develops, and the star descends the DCF, quickly ascending back as substantial mass loss reduces the now CN self-enriched envelope. At this point, the core has ceased nuclear burning and cools under degenerate conditions, with models showing residual luminosity from surface contraction and shell burning for a brief post-DCF period. The final remnant consists of a degenerate He WD ({see Table \ref{tab:IIIWDs}}; centre mass fractions of $^{4}{\rm He} = 0.99$, $^{12}{\rm C} = 3.7\times10^{-5}$, $^{16}{\rm O} = 8.45\times10^{-8}$).

\begin{deluxetable*}{c|c|c|c}
    \label{tab:IIIWDs}
    \tablecaption{{Progenitor and resulting white dwarf masses, and total evolutionary timescales in our Pop III models compared with \citealt{2023MNRAS.525.4700L} (LM23).}}
    \tablehead{
    \colhead{\shortstack{Progenitor Mass \\ (M$_\odot$)}} & 
    \colhead{\shortstack{M$_{\rm WD}$ \\ (M$_\odot$)}} & 
    \colhead{\shortstack{M$_{\rm WD,~LM23}$ \\ (M$_\odot$)}} & 
    \colhead{\shortstack{$\Gamma$ \\ (Gyr)}}}
    \startdata
    0.80  & 0.451 & 0.456 & 12.36 \\
    0.85  & 0.466 & 0.471 & 10.23 \\
    0.90  & 0.544 & 0.494 & 8.59  \\
    0.95  & 0.602 & 0.590 & 7.14  \\
    1.00  & 0.655 & 0.651 & 6.02  \\
    \enddata
\end{deluxetable*}

Earlier on, mass loss at the T-RGB reaches a value of $\vert\dot{\rm M}\vert = 7.6\times10^{-8}~M_\odot~{\rm yr}^{-1}$, and by the T-DCF,  $\vert\dot{\rm M}\vert = 3\times10^{-7}~M_\odot~{\rm yr}^{-1}$. Although radiative winds are expected to be inefficient in metal-free stars, surface enrichment of CNO elements from the shell merger increases envelope opacity and facilitates mass loss through pulsation-enhanced mechanisms. For a 0.8 $M_\odot$ Pop~III model, the cumulative ISM injected mass of carbon was found to be $M^{\rm ej}_{\rm C}(t) = 1.9\times10^{-3}~M_\odot$ by the WD phase, and the total ejected mass of nitrogen was found to be $M^{\rm ej}_{\rm N}(t) = 2.5\times10^{-3}~M_\odot$. The material expelled during the DCF phase contributes modest, yet non-negligible, chemical yields to the interstellar medium; winds enriched in carbon and nitrogen due to elevated surface abundances can seed second-generation stars in low-mass halos, and although these ejecta lack heavier $\alpha-$elements and iron-group species, their CNO-rich composition may leave distinct abundance signatures observable in extremely metal-poor halo stars.

\subsection{$0.85~M_\odot$ Pop~III Stellar Model}\label{sec:085}

Energy generation during the main-sequence (MS) for a $0.85~M_\odot$ Pop~III model is also dominated by the proton-proton (pp) chain. The core remains radiatively stable, whilst the outer layers develop a modest convective envelope, induced by partial ionisation and changes of temperature gradients. 

This convective region endures as the model evolves into the sub-giant (SG) phase. At this point, the hydrogen-burning shell advances outward with the core contracting and heating. As contraction continues and the central temperature rises, helium ignites in an off-centre shell and propagates inward; only after this flash episode does the star settle into a phase of quiescent core helium burning. Throughout the RGB phase, energy generation is initially dominated by the pp-chain, as the models begin essentially devoid of CNO elements. Gradually, however, $3\alpha$ reactions in the core produce $^{12}{\rm C}$, seeding the CNO cycle, which then becomes increasingly significant in regions where hydrogen remains abundant. $0.8~M_\odot$ Pop~III stellar models evolve similarly to the $0.85~M_\odot$ case, with key differences arising in core mass, temperature, and density---particularly beyond the tip of the RGB.

\begin{figure*}[t]
    \centering
    \includegraphics[width = 0.49\linewidth]{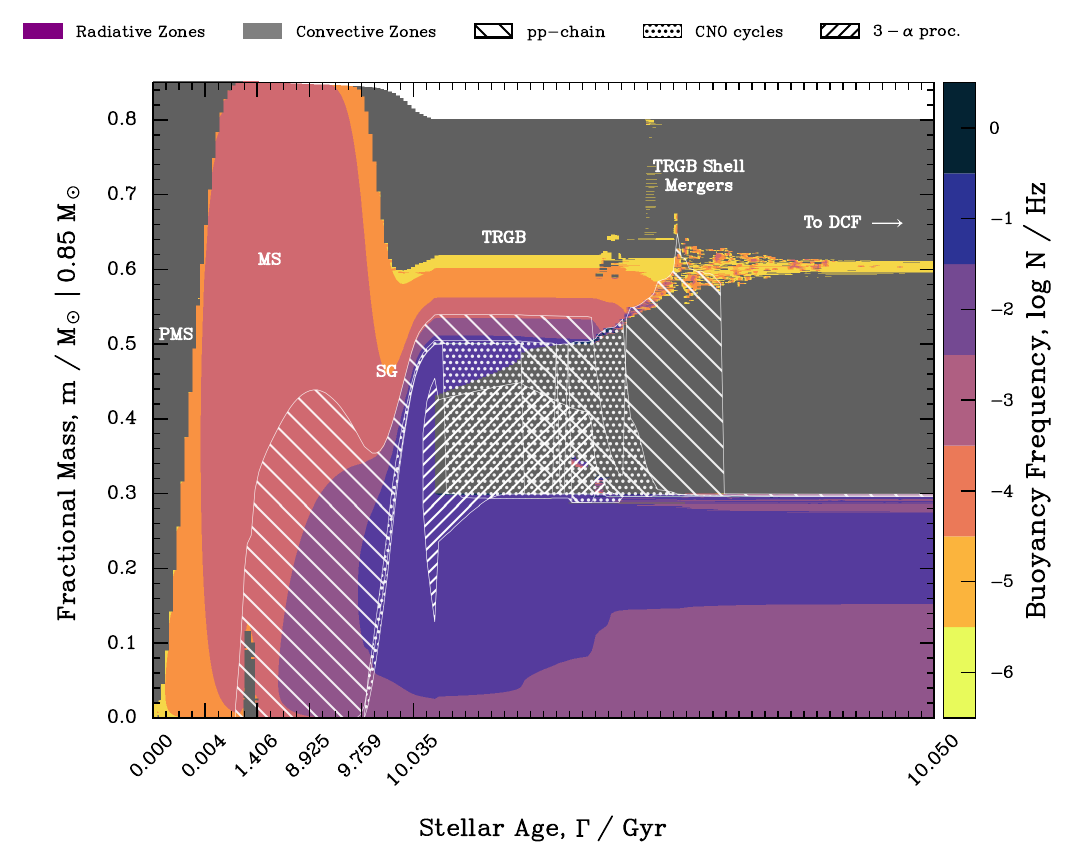}
    \includegraphics[width = 0.49\linewidth]{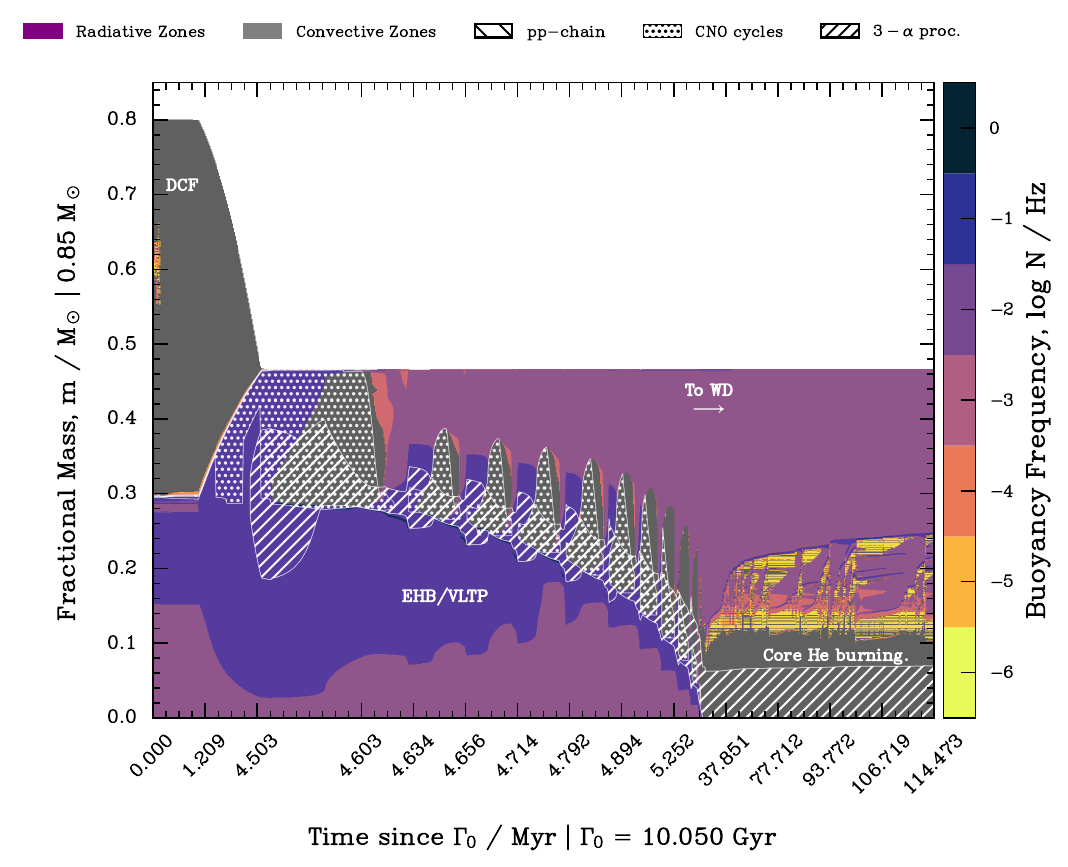}\\
    \caption{Kippenhahn diagrams for a $0.85~M_\odot$ Pop~III model, illustrating the evolutionary stages from the pre-main sequence (PMS) to the T-RGB on the left, and from the DCF to the WD phase on the right. The time axis on the right-hand panel is expressed in millions of years, starting from $10.05$~Gyr, which includes the T-RGB and shell merger episodes. See Figure~\ref{fig:kipp08} for the meaning of the colour scheme and line styles.}
    \label{fig:kipp085}
\end{figure*}

At $\Gamma \approx 10.04$~Gyr, helium flash-driven shell mergers at the tip ot the RGB promote deep convective dredge-up episodes, enriching the surface of the model with CNO-processes material, and leading to surface mass fractions of ${\rm X}_{\rm CNO} \sim 0.013$ (${\rm X}_{^{12}\rm C} \sim 0.006$, ${\rm X}_{^{14}\rm N} \sim 0.007$). Consequently, the model evolves into a nitrogen-rich ultra metal-poor star, with a total mass of $M \sim 0.79~M_\odot$ until mass loss during the DCF (see Figure~\ref{fig:kipp085}). 

{Helium ignition in this case occurs off-centre under degenerate conditions when the model presents $T_{\rm eff} \sim 23~000~K$ and $L \sim 670~L_\odot$. Neutrino cooling and the growth of the compact helium core, driven by substantial mass loss during the T-DCF and hot dredge-up episodes, lead to this ignition profile. The result is a brief extreme horizontal branch (EHB) phase: the helium flash itself burns only a tiny amount of helium, but it triggers the onset of a longer-lived period of steady core helium burning within a thin envelope---a hallmark of helium flash ignition. This phase appears as a two-pulse feature in the H-R diagram (see bottom panel of Figure~\ref{fig:kipp085}): a first high-amplitude excursion in $\log(L)-\log(T_{\rm eff})$ space followed by shorter amplitude phases.}

Following to the T-DCF at $T_{\rm eff} \approx 3300~K$ and $\log(L/L_\odot) \approx 3.2$, the model undergoes a brief contraction phase accompanied by moderate heating, resulting in a mild shift on the H–R diagram to $T_{\rm eff} \approx 3500~K$ and $\log(L/L_\odot) \approx 3.0$. This transition is accompanied by the development of a small radiative region beneath the convective envelope (in mass coordinates) and a velocity inversion at the outer structural boundary, before entering the planetary nebula (PN) phase. The pre-WD structure is characterised by a degenerate cooling core and an extended radiative envelope, with an envelope mass of $\approx0.0012~M_\odot$, a helium core mass of $\approx0.464~M_\odot$, a carbon-oxygen core mass of $\approx 0.162~M_\odot$, and negligible masses for the O-Ne, Fe, and neutron-rich cores. At this phase, the model reaches about $T_{\rm eff} \sim 51~000~K$, $L\sim350~L_\odot$ and $\log(g) \sim 4.3$ dex.

During early post-DCF evolution, $\alpha-$capture on residual nitrogen proceeds briefly via the sequence $^{14}{\rm N}(\alpha,\gamma)^{18}{\rm F}(\beta^+\nu_e)^{18}{\rm O}(\alpha,\gamma)^{22}{\rm Ne}$, producing neutrinos that dominate the energy loss during this stage (e.g., \citealt{2020ApJ...893..133F}). This reaction chain operates over short timescales ($\lesssim 10^{3}$ years) and does not contribute significantly to long-term energy generation. Its main legacy is compositional: nearly all $^{14}{\rm N}$ is transmuted into $^{22}{\rm Ne}$, yielding considerable core mass fractions \citep{2021ApJ...910...24C}. Such enrichments may even impact the asteroseismology of the resulting WDs, shifting g-mode periods by up to 1\%. After this brief convective and nuclear episode, the star transitions into a fully radiative WD precursor on thermal timescales of $10^{4-5}$ years. 

In our models, the application of Reimers and Bl{\"o}cker-type mass-loss prescriptions leads to efficient stripping of the outer envelope, enabling the emergence of EHB-like structures---consistent with prior works (e.g., \citealt{2023MNRAS.525.4700L}). Mass loss rates reach maximum values of $\vert\dot{\rm M}\vert = 6\times10^{-8}~M_\odot~{\rm yr}^{-1}$ at the T-RGB, and $\vert\dot{\rm M}\vert = 3.2\times10^{-7}~M_\odot~{\rm yr}^{-1}$ at the T-DCF. Such EHB stars are low-mass, helium-core burning objects with extremely thin hydrogen envelopes, typically manifesting observationally as O/B-type sub-dwarfs (sdO/sdB; \citealt{1996ApJ...466..359D}). In a post-DCF context, such EHB phases may be triggered by reigniting residual helium in the envelope. These phenomena, on one hand, may often be attributed to the $\epsilon-$(epsilon)mechanism\footnote{Unlike the $\kappa-$(kappa)mechanism, which operates in the stellar envelope through opacity variations, the $\epsilon-$mechanism arises in regions where nuclear burning is still active, such as residual hydrogen or helium-burning layers. Since the nuclear reaction rates are highly sensitive to temperature ($\epsilon \propto T^\nu$), small perturbations can lead to significant energy fluctuations, driving pulsations if the energy input exceeds damping effects. Whilst this mechanism is typically more relevant in massive stars with active core burning (e.g., \citealt{2018A&A...614A.136B}, and references therein), its influence in hot sub-dwarfs may contribute to high-frequency oscillations, particularly in evolutionary phases where thin burning shells persist as has been modelled for low-mass Pop~III stars.} \citep{1989nos..book.....U}, whereby variations in nuclear burning drive high-frequency oscillations. Our models show that residual burning layers in low-mass Pop~III stars can also activate this process, contributing to brief pulsational episodes.

\begin{figure}[t]
    \centering
    \includegraphics[width = \linewidth]{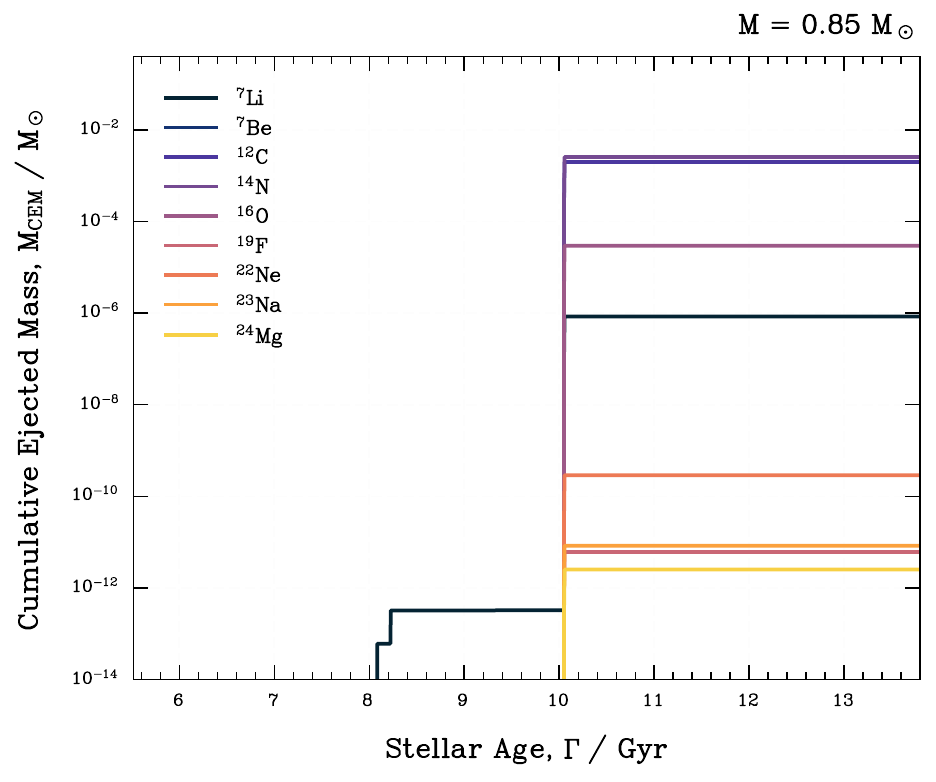}\\
    \includegraphics[width = \linewidth]{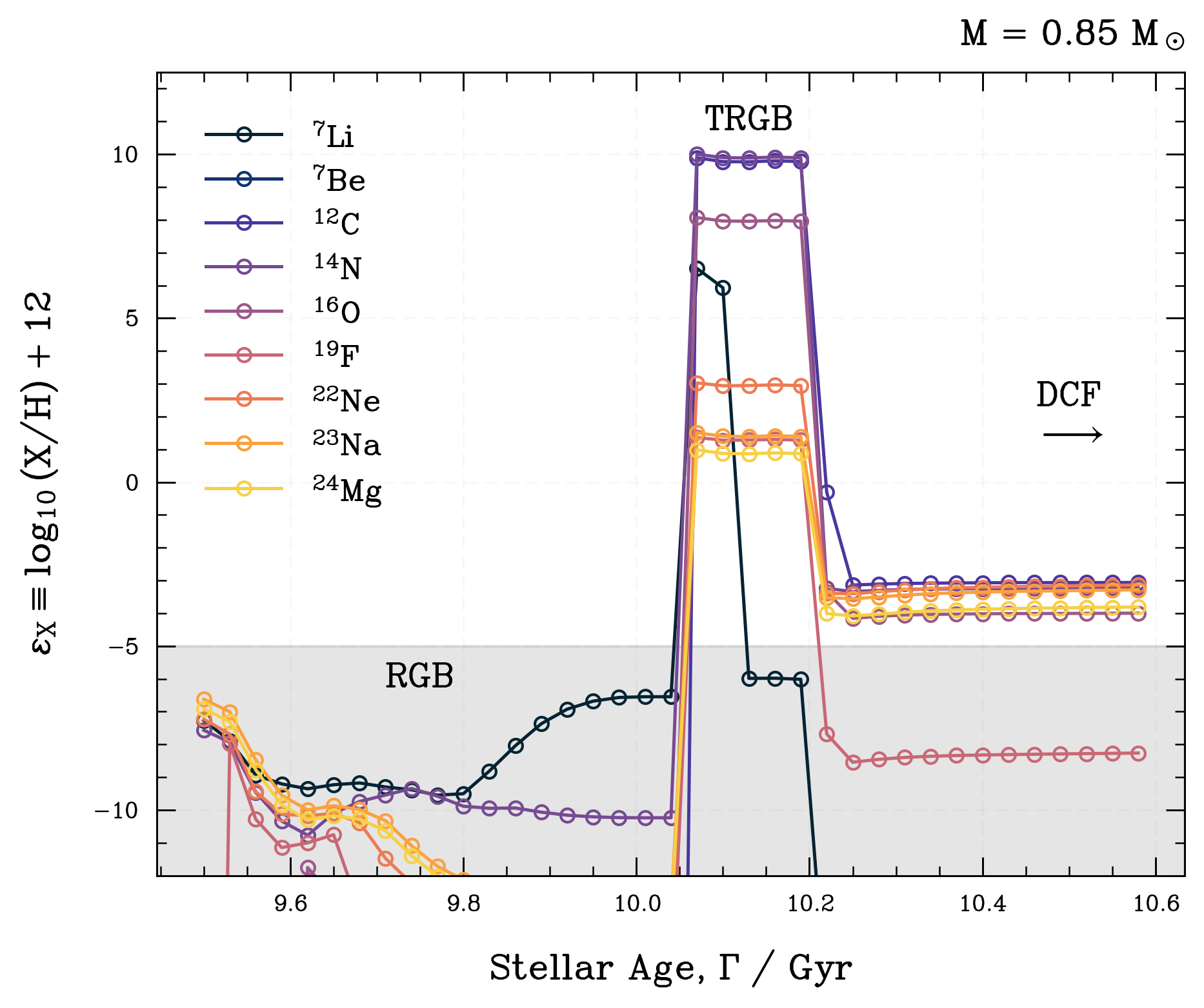}
    \caption{{\it First panel}: Cumulative ejected mass of surface Li, Be and heavier elements (C, N, O, F, Ne, Na, and Mg) as a function of stellar age for a 0.85 $M_\odot$ Pop~III star. {\it Second panel}: Surface abundance evolution of synthesised elements in a 0.85 $M_\odot$ Pop~III stellar model, as a function of stellar age during the shell mergers episodes (see text). Abundances are expressed as $\epsilon_X \equiv \log_{10}(X/\mathrm{H}) + 12$. The grey-shaded region indicates the typical spectroscopic detection limit for current instruments.}
    \label{fig:085ejected}
\end{figure}

Once nuclear burning ceases, the star becomes a CO WD with a final mass of $M_{\rm WD} \sim 0.466~M_\odot$ (centre mass fractions of $^{12}{\rm C} = 0.251$ and $^{16}{\rm O} = 0.741$). From a population synthesis standpoint, this evolutionary track contributes to the chemical enrichment of the early interstellar medium (ISM) through selective CNO dredge-up, albeit with low total yield due to the weak winds and absence of Type II SN ejecta. Whilst standard models predict that old stellar populations emit primarily in the IR-optical range, EHB stars can introduce a substantial UV component to their host’s spectral energy distribution (SED), particularly in globular clusters and elliptical galaxies (e.g., \citealt{1994AJ....107.1408L, 2024MNRAS.529..628V}). This excess UV flux, if present, may serve as a fossil indicator of early chemical enrichment by metal-free or extremely metal-poor stars. Figure~\ref{fig:085ejected} presents the cumulative ISM injected mass of carbon for 0.85 $M_\odot$ Pop~III star, which reaches a value of $M^{\rm ej}_{\rm C}(t) = 2\times10^{-3}~M_\odot$ by the WD phase, and the total ejected mass of nitrogen, the second most abundant ejected element followed by oxygen and lithium (albeit on orders of magnitude smaller) that was found to be $M^{\rm ej}_{\rm N}(t) = 2.6\times10^{-3}~M_\odot$. For 0.9 $M_\odot$ Pop~III stars, the cumulative ISM injected mass of carbon was found to be $M^{\rm ej}_{\rm C}(t) = 1.7\times10^{-3}~M_\odot$, whilst nitrogen accounts for $M^{\rm ej}_{\rm N}(t) = 2.15\times10^{-3}~M_\odot$, in comparison.

In Figure~\ref{fig:085ejected}, we also present the temporal evolution of surface chemical abundances, expressed as logarithmic abundance ratios $\epsilon_X \equiv \log_{10}(X/\mathrm{H}) + 12$, for key elements synthesised in a 0.85 $M_\odot$ Pop~III stellar model. The figure illustrates how surface enrichment episodes---primarily during the T-RGB phase---lead to temporary increases in the abundances of elements such as carbon ($^{12}$C), nitrogen ($^{14}$N), oxygen ($^{16}$O), and fluorine ($^{19}$F). These enhancements result from convective dredge-up processes and envelope instabilities that transport freshly synthesised material to the stellar surface; however, once the star enters the WD phase and mass loss subsides, the surface abundances of all elements drop to levels generally below $\epsilon_X \lesssim 2$. {Our models indicate that low-mass Pop III stars experience only a brief period after leaving the main sequence, in which convective processes bring heavy elements to the surface at levels approaching detectability, and outside this short-lived phase, surface abundances remain extremely low, with $\epsilon_X \lesssim -5$, as gravitational settling and inefficient mass loss suppress any observable enrichment (e.g., \citealt{2024A&A...691A.245C, 2025arXiv250921643J}). Consequently, for the vast majority of their lifetimes---including both the main-sequence and late cooling phases---these stars would remain effectively invisible to spectroscopic surveys, and direct detection through traditional spectroscopy is therefore feasible only under rare, transient conditions, highlighting the necessity of alternative observational approaches to characterise the population.}

\subsection{$0.9~M_\odot$ Stellar Model}\label{sec:09}

\begin{figure*}[t]
    \centering
    \includegraphics[width = \linewidth]{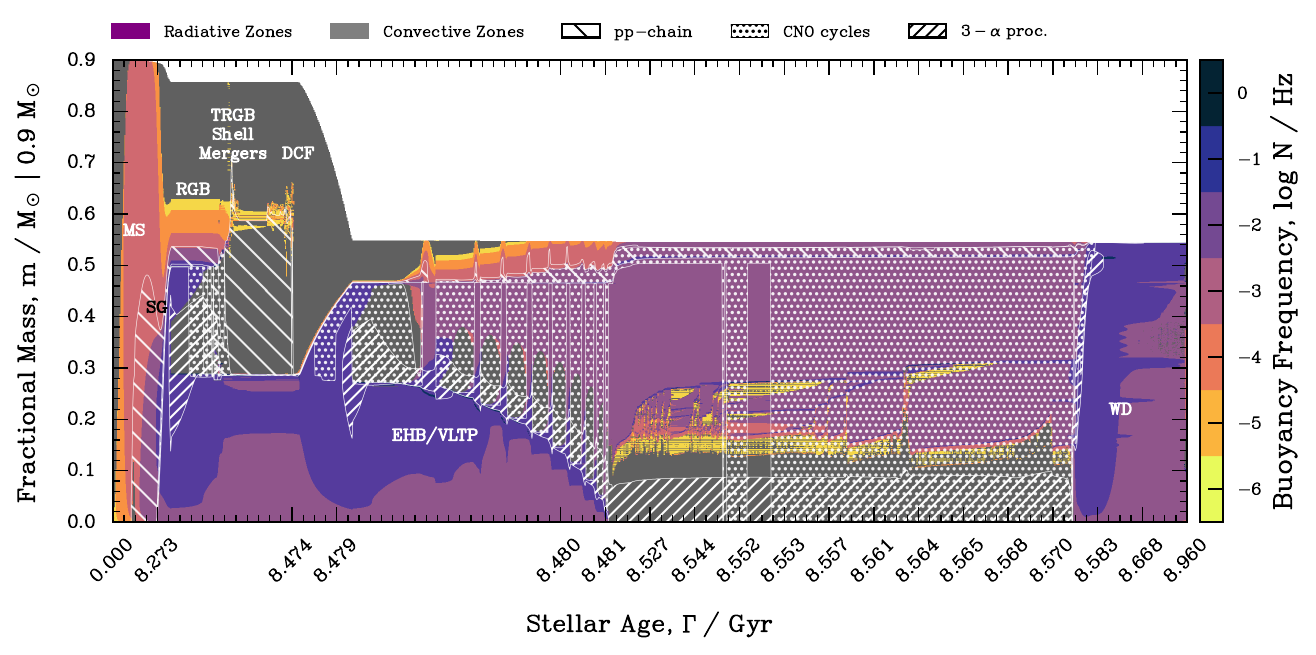}
    \caption{Kippenhahn diagram for a $0.9~M_\odot$ Pop~III star, shown from the PMS to the WD phase.}
    \label{fig:kipp09}
\end{figure*}

A 0.9 $M_\odot$ Pop~III star follows a similar morphological trajectory compared to a 0.85 $M_\odot$ Pop~III star, albeit with further complexity post-DCF (see Figure~\ref{fig:kipp09}). At $\Gamma \approx 8.4$~Gyr, shell mergers at the base of the DCF trigger efficient dredge-up, producing surface abundances of ${\rm X}_{\rm CNO} \sim 0.0125$ (${\rm X}_{^{12}\rm C} \sim 0.0055$, ${\rm X}_{^{14}\rm N} \sim 0.0069$), again resulting in a nitrogen-rich low-mass star ($M \sim 0.79~M_\odot$).

After the T-DCF and associated envelope loss, the star begins contracting, tracing a downward path on the H-R diagram. Between $T_{\rm eff} \sim 4200-15~000~K$, $L \sim 100-200~L_\odot$, and $\log(g) = 1.1-3.7$ dex, helium shell flashes are triggered due to helium accumulation atop a partially degenerate CO core ($\eta_c = 20$). These instabilities occur because the degenerate shell cannot expand to relieve pressure as temperature rises, resulting in rapid, unstable helium ignition via the $3\alpha$ process. Brief EHB-like phases follow, but the low residual envelope mass prevents prolonged helium fusion. The star quickly exhausts its fuel and transitions to a CO WD with $M_{\rm WD} \sim 0.543~M_\odot$ (centre mass fractions of $^{12}{\rm C} = 0.32$ and $^{16}{\rm O} = 0.672$).

These models suggest that EHB-like behaviour may {\it naturally} emerge in zero-metallicity stars under conditions of intense, parametrised mass loss and core degeneracy---similar to such pathways that have been primarily explored in metal-rich populations (e.g., \citealt{2008MmSAI..79..375H, 2018A&A...614A.136B}). The brevity of the EHB phase (e.g., $\approx$60~Myr; Figure~\ref{fig:kipp085}), however, demonstrate its rarity, yet it may offer a pathway to UV-bright signatures in ancient stellar populations, which---although not definitively explored---could represent indirect tracers of Pop~III enrichment.

\subsection{$0.95-1.0~M_\odot$ Stellar Models}\label{sec:095-1}

As metal-free low-mass stars, $0.8-1~M_\odot$ Pop~III stars spend a prolonged portion of their lifetimes on the MS, primarily powered by hydrogen fusion via the pp-chain. The lack of metals results in relatively lower opacities, influencing the star's internal structures and diminishing its stable core hydrogen-burning phase, sustaining stable core hydrogen burning for approximately $65-85\%$ of their total lifetimes (as computed up to the WD phase). The reduced opacity of metal-free gas leads to more compact and hotter cores compared to metal-rich counterparts, which alters the efficiency of energy transport and slightly shortens MS lifetimes relative to what would be expected for stars of similar mass but higher metallicity.

Mixing is triggered during shell hydrogen-burning phases when the inversion of the molecular weight gradient drives mixing between the core and the envelope \citep{2004sipp.book.....H, 2013sse..book.....K}. This process brings $^{14}$N and $^{12}$C from the core to the surface (dredge-up events), creating potentially observable anomalies that could be adopted to distinguish Pop~III stars from their metal-rich counterparts. The timescale and efficiency of mixing vary with mass and evolutionary phase, typically occurring after $6-10$~Gyr, influencing surface abundance patterns more significantly in stars with masses near the upper end of the low-mass range.

\begin{figure*}[t]
    \centering
    \includegraphics[width = \linewidth]{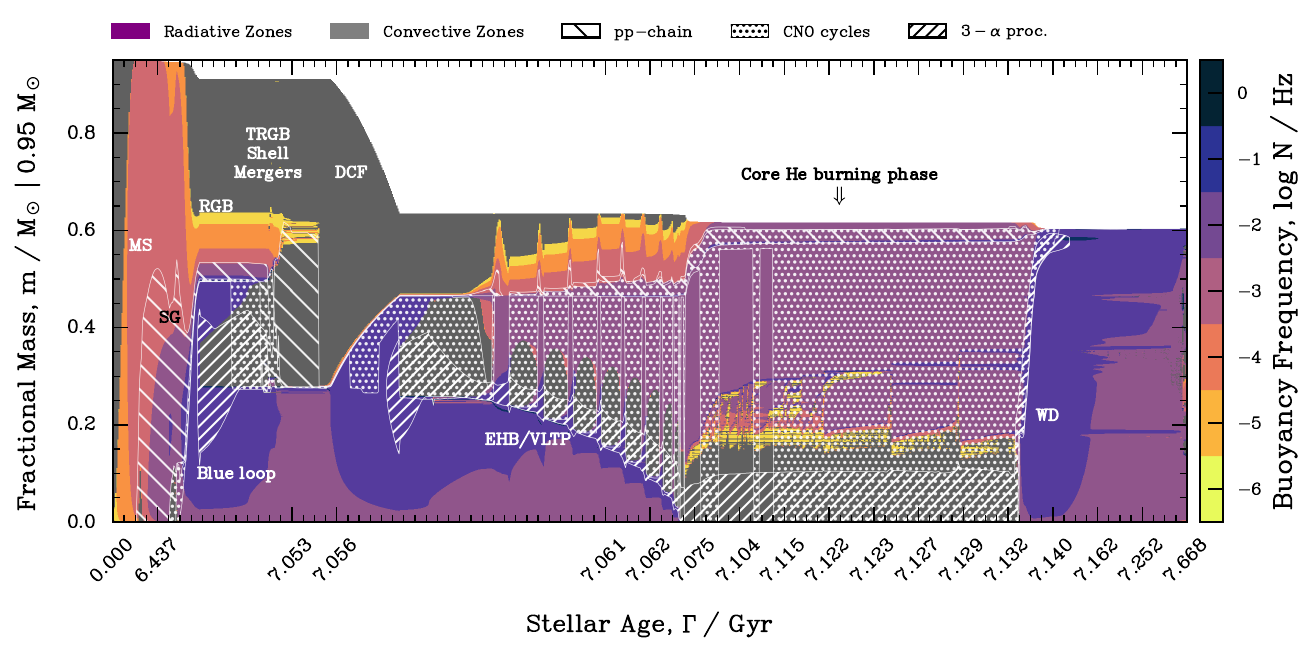}
    \includegraphics[width = \linewidth]{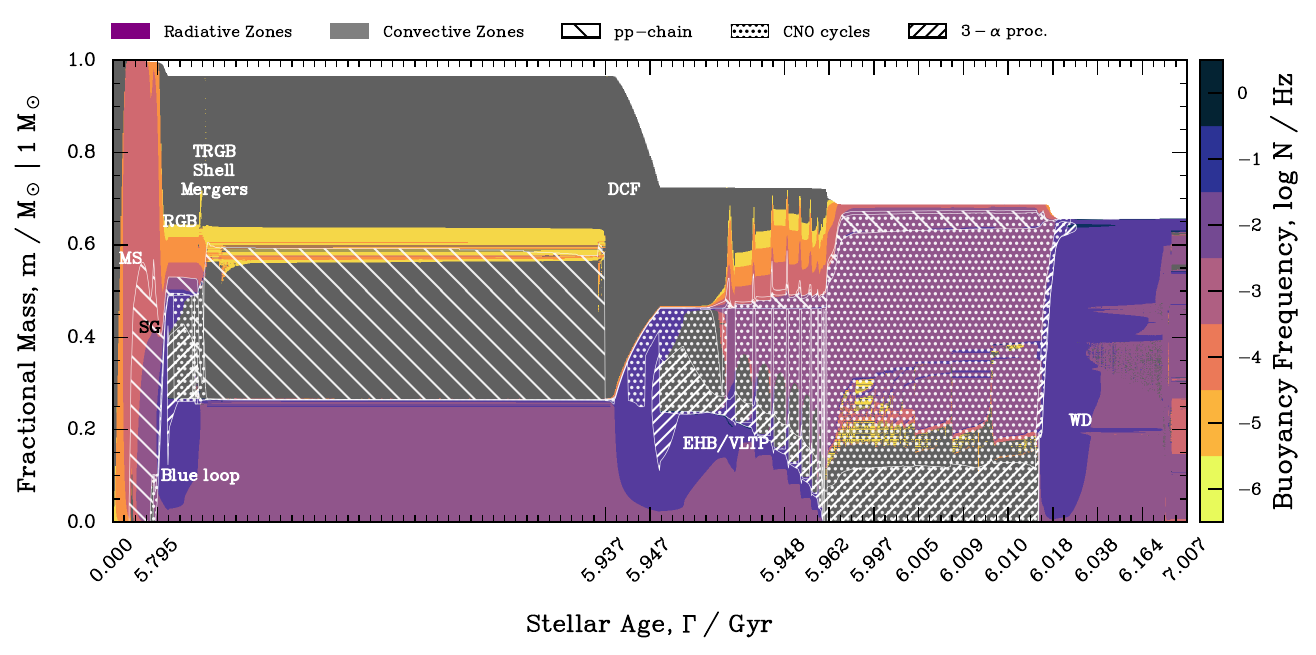}
    \caption{Kippenhahn diagrams for $0.95$ and $1~M_\odot$ Pop~III models, shown from the PMS to the WD phase.}
    \label{fig:kipp095-1}
\end{figure*}

For $0.95~M_\odot$ and $1~M_\odot$ (Figure~\ref{fig:kipp095-1}), however, the sub-giant phase differs from their metal-free lower-mass pairs slightly due to an extremely short-lived blueward evolutionary loop that lifts central hydrogen abundances to up to $X_c = 0.0016$ (0.95 $M_\odot$) and $X_c = 0.0017$ (1 $M_\odot$)---lasting only $\sim$9 and $\sim$9.2 years, respectively, and initiating at $\Gamma \sim 6.850$ and $5.758$~Gyr---triggered by the onset of $3\alpha$ helium ignition in the core, a consequence of the competition between core helium burning and shell hydrogen burning. Once helium burning stabilises and dominates the energy production, the envelope expands again, pushing the star back up the RGB. For stars at the lower end of this mass range ($0.8-0.9~M_\odot$), core convection during the main sequence is minimal or absent, with energy generation dominated by the pp-chain and a negligible contribution from the CNO cycle due to extremely low initial abundances of CNO elements. These stars exhibit prolonged stable hydrogen burning without significant core convection, contrasting with higher-mass cases where a brief convective core may develop during sub-giant evolution.

As with the lower-masses cases, shell mergers at the base of the DCF phase dredge CNO elements to the surface of such stars at levels ${\rm X}_{\rm CNO}\sim0.0116$ for $0.95~M_\odot$ stars (${\rm X}_{^{12}\rm C} \sim 0.0052$ and ${\rm X}_{^{14}\rm N} \sim 0.0063$) and ${\rm X}_{\rm CNO}\sim 0.0107$ for $1~M_\odot$ stars (${\rm X}_{^{12}\rm C} \sim 0.0049$ and ${\rm X}_{^{14}\rm N} \sim 0.0057$). These become nitrogen-rich $0.63~M_\odot$ and $0.72~M_\odot$ stars at these phases. The stars then descend the H-R diagram, featuring helium shell flashes by reaching the T-DCF followed by brief, very low-amplitude EHB-like pulsations at $T_{\rm eff} \sim 28480~K$, $L\sim70~L_\odot$, and $\log(g) = 5.14$ dex for $0.95~M_\odot$ stars, and $T_{\rm eff} \sim 27520~K$, $L\sim125~L_\odot$, and $\log(g) = 4.89$ dex for $1~M_\odot$ stars.

At these phases, the star exhibits a layered internal structure beginning with a convective core surrounded by a tiny radiative shell whose energy generation is primarily via $3\alpha$ processes, followed by a convective zone, and an outer envelope featuring an inner radiative zone (CNO-dominated) surrounded by a convective outer layer whose energy generation is mainly through the pp-chain. Following the end of core-helium burning, the convective core ceases, the star develops deep radiative zones within its interior and energy generation is shifted towards the outer envelope; at the same time, the core cools down, leading to a radial expansion of the outer envelope and a strong incursion towards the red-side of the H-R diagram up to $T_{\rm eff} \sim 8500~K$ and $L \sim 2000~L_\odot$, and $\log(g) = 1.60$ dex at $7.13$~Gyr for $0.95~M_\odot$ stars, and $T_{\rm eff} \sim 5000~K$ and $L \sim 1700~L_\odot$, and $\log(g) = 0.7$ dex at $6.01$~Gyr for $1~M_\odot$ stars, resembling red or yellow supergiant (RSG/YSG) stars of K-A types {though their luminosities are $5–50$ times lower than those} (e.g., \citealt{1971MNRAS.152..121P, 1992A&A...259..600A, 1998MNRAS.298..525P, 2000ApJ...528..368H, 2001ApJ...560..934S, 2009ApJ...703..441D, 2011BSRSL..80..266M, 2012A&A...537A.146E, 2015A&A...575A..60M}).

Unlike lower-mass cases, $0.95~M_\odot$ and $1~M_\odot$ Pop~III stars experience stronger mass loss during the RGB phase, reaching rates of $|\dot{M}| = 2\times10^{-7}~M_\odot~\text{yr}^{-1}$ for both masses. This is followed by peaks at T-DCF, with $|\dot{M}| = 4.4\times10^{-8}~M_\odot~\text{yr}^{-1}$ for the $0.95~M_\odot$ star and $|\dot{M}| = 7.1\times10^{-8}~M_\odot~\text{yr}^{-1}$ for the $1~M_\odot$ case. The cumulative mass of carbon expelled into the ISM was found to be $M^{\rm ej}_{\rm C}(t)_{0.95~M_\odot} = 1.6\times10^{-3}~M_\odot$ and $M^{\rm ej}_{\rm C}(t)_{1~M_\odot} = 1.47\times10^{-3}~M_\odot$, whilst for nitrogen it was $M^{\rm ej}_{\rm N}(t)_{0.95~M_\odot} = 1.95\times10^{-3}~M_\odot$ and $M^{\rm ej}_{\rm N}(t)_{1~M_\odot} = 1.83\times10^{-3}~M_\odot$. Although modest, these RGB winds contribute non-negligibly to the chemical enrichment of the ISM, especially given the elevated surface abundances of carbon and nitrogen. Though devoid of heavier $\alpha$ elements and iron-group species, it may also imprint distinctive chemical signatures in extremely metal-poor second-generation halo stars forming in low-mass mini-halos.

As further evolution proceeds, core-cooling continues and the inner radiative layers diminish up to $(T_{\rm eff}/{\rm K}, L/L_\odot,\log(g)) \approx (130000, 960, 6.65)$ for $0.95~M_\odot$ stars and $(T_{\rm eff}/{\rm K}, L/L_\odot,\log(g)) \approx (150000, 1500, 6.75)$ for $1~M_\odot$ Pop~III stars resembling blue giants of OB-types (e.g., \citealt{1984PhR...105..329I, 2005ApJ...627..477M, 2008A&A...489..713L}) before settling into the CO WD sequences with masses $M_{\rm {\rm WD},~0.95~M_\odot prog.} \approx 0.6~M_\odot$ (centre mass fractions of $^{12}{\rm C} = 0.345$ and $^{16}{\rm O} = 0.646$) and $M_{\rm {\rm WD},~1~M_\odot prog.} \approx 0.65~M_\odot$ (centre mass fractions of $^{12}{\rm C} = 0.326$ and $^{16}{\rm O} = 0.662$). It may also be notable that the early PMS and ZAMS phases exhibit relatively rapid contraction and heating, with the onset of hydrogen burning occurring slightly earlier than in metal-rich stars due to lower opacity, influencing the initial convective structure and early mixing processes.

\end{document}